\documentclass[12pt]{article}
\pdfoutput=1
\usepackage{amssymb, amsmath,amsfonts}
\usepackage{multirow}
\usepackage{mathrsfs}
\usepackage{array}
\usepackage{cite}
\usepackage{booktabs}
\usepackage{tikz}
\usepackage{pgfplots}
\usepackage{float}
\usepackage{subfigure, graphicx}
\usepackage[pdftex, bookmarks=true,colorlinks,linkcolor=red,urlcolor=blue,citecolor=blue]{hyperref}
\usepackage{cite}
\usepackage{slashed}
\usepackage{tabularx}

\usepackage{tikz}

\makeatletter\@addtoreset{equation}{section}\makeatother

\def\be{\begin{equation}}
\def\ee{\end{equation}}
\def\bea{\begin{eqnarray}}
\def\eea{\end{eqnarray}}

\newcommand{\nn}{\nonumber}

\def\Dslash{\,\,{\raise.15ex\hbox{/}\mkern-12mu D}}
\def\Dbarslash{\,\,{\raise.15ex\hbox{/}\mkern-12mu {\bar D}}}
\def\delslash{\,\,{\raise.15ex\hbox{/}\mkern-9mu \partial}}
\def\delbarslash{\,\,{\raise.15ex\hbox{/}\mkern-9mu {\bar\partial}}}
\def\pslash{\,\,{\raise.15ex\hbox{/}\mkern-9mu p}}
\def\calDslash{\,\,{\raise.15ex\hbox{/}\mkern-12mu {\cal D}}}

\makeatletter\@addtoreset{equation}{section}\makeatother

\renewcommand{\title}[1]{\vbox{\center\LARGE{#1}}\vspace{5mm}}
\renewcommand{\author}[1]{\vbox{\center#1}\vspace{5mm}}
\newcommand{\address}[1]{\vbox{\center\em#1}}

\def\arXiv#1{\href{http://arxiv.org/abs/#1}{arXiv:#1}}
\def\arXiv#1#2{\href{http://arxiv.org/abs/#1}{arXiv:#1}}

\def\doi#1#2{\href{http://doi.org/#1}{#2}}

\begin{document}

\unitlength = .5mm

\begin{titlepage}
\vspace{.5cm}
 
\begin{center}
\hfill \\

\title{Holographic study of shear viscosity and butterfly velocity for magnetic field-driven quantum criticality}
\vskip 0.5cm

{Jun-Kun Zhao$^{\,a}$}\footnote{Email: {\tt junkunzhao@itp.ac.cn}} and {Li Li$^{\,a,b,c}$}\footnote{Email: {\tt liliphy@itp.ac.cn}}

\address{${}^{a}$Institute of Theoretical Physics, Chinese Academy of Sciences, Beijing 100190, China}
\address{${}^{b}$School of Physical Sciences, University of Chinese Academy of Sciences, No.19A Yuquan Road, Beijing 100049, China}
\address{${}^{c}$School of Fundamental Physics and Mathematical Sciences, Hangzhou Institute for Advanced Study, UCAS, Hangzhou 310024, China}
\end{center}
\vskip .8cm

\abstract{
We investigate the shear viscosity and butterfly velocity of a magnetic field-induced quantum phase transition in five dimensional Einstein-Maxwell-Chern-Simons theory, which is holographically dual to a class of strongly coupled quantum field theories with chiral anomalies. Our analysis reveals that the ratio of longitudinal shear viscosity to entropy density $\eta_\parallel/s$ exhibits a pronounced non-monotonic dependence on temperature $T$  when the magnetic field $B$ is slightly below the critical value $B_c$ of the quantum phase transition. In particular, it can develop a distinct minimum at an intermediate temperature. This contrasts sharply with the monotonic temperature scaling observed at and above $B_c$, where $\eta_\parallel/s$ follows the scaling $T^{2/3}$ at $B=B_c$ and transitions to $T^2$ for $B>B_c$ as $T\to0$. The non-vanishing of $\eta_\parallel/s$ for $B<B_c$ in the zero temperature limit suggests that it could serve as a good order parameter of the quantum phase transition. We also find that all butterfly velocities change dramatically near the quantum phase transition, and thus their derivatives with respect to $B$ can be independently used to detect the quantum critical point.
}
\vfill

\end{titlepage}

\begingroup 
\hypersetup{linkcolor=black}
\tableofcontents
\endgroup




\section{Introduction}
Quantum phase transition (QPT), driven by quantum fluctuations at absolute zero temperature, plays a crucial role in understanding exotic properties of strongly-correlated quantum matter~\cite{sachdev1999quantum}.
Specifically, the quantum critical point (QCP) can dominate a broad regime of the phase diagram away from it, offering valuable insights into the mysteries, for example, in high-temperature superconductors and strangle metals. Moreover, the dynamics near the QCP may go beyond the traditional framework of the Landau-Ginzburg-Wilson paradigm~\cite{Senthil:2003eed,Senthil:2023vqd,Armitage:2017cjs,Baggioli:2021xuv}. A hallmark of quantum criticality is the emergence of scale invariance at the QCP, where the system could be governed by conformal field theory (CFT). This suggests a profound connection to the Anti-de Sitter/Conformal Field Theory (AdS/CFT) correspondence~\cite{Maldacena:1997re}, particularly when the CFT possesses a large central charge. Indeed, holographic techniques have been applied to describe strongly coupled quantum critical systems without quasi-particles, providing a non-perturbative and readily accessible approach for investigating various transports as well as the far-from-equilibrium dynamics~\cite{Hartnollbook,Zaanen:2015oix,Liu:2018crr}. In holography, significant progress has been made in modeling QPTs in dual strongly coupled many-body systems (see \emph{e.g.}~\cite{Iqbal:2010eh,DHoker:2010onp,Hartnoll:2011pp,Donos:2012js,Landsteiner:2015pdh}). Among them, the most intriguing one is the QPT driven by a magnetic field, which occurs without symmetry breaking and is described by the Einstein-Maxwell-Chern-Simons theory (EMCS)~\cite{DHoker:2010onp}.

The five dimensional EMCS theory is one of the benchmark models in AdS/CFT, which is dual to a class of strongly coupled systems with chiral anomaly~\cite{Witten:1998qj}. It offers a robust testing ground in the construction of chiral hydrodynamic constitutive relations with magnetic field~\cite{Ammon:2020rvg} and the study of anomalous hydrodynamics effective action~\cite{Rangamani:2023mok}. In this theory, the interplay between finite charge density and magnetic field yields a strongly coupled anisotropic system with a magnetic field induced QPT~\cite{DHoker:2009mmn,DHoker:2010zpp,DHoker:2010onp,DHoker:2012rlj}. Although it has received a lot of interest~\cite{Astefanesei:2010dk,Ammon:2016szz,Ammon:2017ded,Cai:2024tyv}, the physical interpretation of this phenomenon remains obscure due to the lack of a good order parameter, and more properties associated with quantum criticality are still undiscovered. Therefore, it is necessary to study it from multiple perspectives. In the present work, we aim at probing the quantum criticality of the QPT by using shear viscosity and butterfly velocity. We note that the shear viscosity in this holographic model was initially computed in~\cite{Ammon:2020rvg}, working at a relatively high temperature and strong magnetic field. Some properties of butterfly velocities were considered in~\cite{Abbasi:2019rhy,Abbasi:2023myj}, which suggested that one must employ at least two butterfly velocities to identify the location of the QCP. Despite this, the fundamental impacts of QPT on shear viscosity and butterfly velocity remain largely unexplored. This motivates us to revisit the question of the properties of shear viscosity and butterfly velocity, focusing on the effects of quantum criticality.

One of the most quantitative applications of holography to strongly correlated systems is the ratio of shear viscosity to entropy density $\eta/s=\hbar/(4\pi k_B)$ that is universal in a large class of theories, known as the Kovtun-Son-Starinets (KSS) bound~\cite{Kovtun:2004de,Kovtun:2003wp}. Nevertheless, there are some ways for which the KSS bound could be violated. Firstly, certain higher derivative corrections to the low-energy Einstein action can push $\eta/s$ below the KSS bound, while causality constraints in the boundary theory impose a revised lower bound~\cite{Brigante:2007nu,Brigante:2008gz,Kats:2007mq}\footnote{For more research on higher derivative corrections, one can refer to, for instance,~\cite{Cai:2008ph,Myers:2010jv,Feng:2015oea,Wang:2016vmm,Buchel:2024umq}. Additionally, a comprehensive review can be found in~\cite{Cremonini:2011iq}.}. 
Second, within Einstein gravity, violations were realized by breaking spacetime symmetries (translations and/or rotations)~\cite{Rebhan:2011vd,Mamo:2012sy,Jain:2014vka,Critelli:2014kra,Landsteiner:2016stv,Finazzo:2016mhm,Giataganas:2017koz,Gursoy:2020kjd}. However, the breaking of rotations itself does not necessarily imply a violation of the KSS bound~\cite{Baggioli:2023yvc,Erdmenger:2011tj}. Moreover, KSS bound can be violated in far-from-equilibrium processes~\cite{Baggioli:2021tzr,Wondrak:2020tzt,Wang:2024nmn}. Another crucial aspect involves the construction of holographic models in which $\eta/s$ exhibits a temperature-dependent behavior, reaching a characteristic minimum at the phase transition--a phenomenon observed for various fluids in nature. However, only a limited number of holographic models incorporating higher-derivative corrections display a non-monotonic temperature dependence $\eta/s$~\cite{Cremonini:2011ej,Cremonini:2012ny} where the effects are perturbatively small. Such a temperature dependence of shear viscosity and the associated minimum at an intermediate temperature are expected and observed in various systems, including classical fluids~\cite{Trachenko:2019ghg}\footnote{
For classical fluids, the occurrence of a minimum in shear viscosity can be understood as follows~\cite{Trachenko:2019ghg}. In the high-temperature gas regime, the shear viscosity scales as $\eta\propto \sqrt{T}$, resulting in an increase viscosity with temperature. In contrast, in the low-temperature liquid regime, the shear viscosity decreases with temperature as $\eta\propto e^{U/T}$ where $U$ is the activation energy.} and strongly coupled dusty plasmas~\cite{2022PhRvR...4c3064H,2023PhRvR...5a3149H}. Moreover, a minimum in $\eta/s$ at QCD transition temperature is expected in quark–gluon plasma~\cite{Bernhard:2019bmu,Adams:2012th}.

On the other hand, it has been revealed that there is an inextricable connection between late-time hydrodynamic transports and early-time chaotic properties in quantum many-body systems~\cite{Gu:2016oyy,Davison:2016ngz,Blake:2016wvh,Grozdanov:2017ajz}. It is well-known that chaos is a fundamental property inherent in thermal systems, manifesting universally across a wide range of physical phenomena. In quantum many-body systems, chaotic dynamics can be rigorously characterized through the exponential growth of the out-of-time-order correlator (OTOC). In a large class of many-body systems, it has been observed  that~\cite{Larkin:1969,Shenker:2013pqa,Roberts:2014isa,Maldacena:2015waa,Jahnke:2018off}
\bea
\langle V(t,\vec{x})\, W(0)\, V(t,\vec{x})\, W(0) \rangle_\beta = 1- \epsilon e^{\lambda \left( t-\frac{|\vec{x}|}{v_B} \right)} +\cdots \,, \label{eq:otoc}
\eea
where $V$ and $W$ denote generic local few-body operators, $\beta=1/T$ represents the inverse temperature, and $\epsilon \sim \mathcal{N}^{-1}$ parameterizes the degrees of freedom $\mathcal{N}$ of the system. Furthermore, $\lambda$ corresponds to the quantum Lyapunov exponent that is bounded by $\lambda\leq 2\pi\beta$~\cite{Maldacena:2015waa}, and $v_ B$ is the butterfly velocity characterizing information scrambling in spatial dimensions. For strongly coupled quantum matter with gravity dual, the quantum chaos can be determined from the two sided eternal black hole dual to the thermofield double state of two identical CFTs $L$ and $R$~\cite{Shenker:2013pqa,Roberts:2014isa}.  Remarkably, the pole-skipping phenomenon clearly demonstrates the connection between quantum chaos and hydrodynamic transports with $\lambda$ and $v_B$ being identified at the pole-skipping point~\cite{Grozdanov:2017ajz,Blake:2017ris,Blake:2018leo}.

In this paper, we study the shear viscosity within the framework of EMCS theory exhibiting a magnetic field-driven QPT. The presence of magnetic field breaks the spatial $SO(3)$ rotational symmetry into $SO(2)$ symmetry in the plane orthogonal to the magnetic field. Consequently, there is one universal transverse shear viscosity $\eta_\perp$, one non-universal longitudinal shear viscosity $\eta_\parallel$, as well as two Hall viscosities. The expression for transverse shear viscosity can be derived analytically and equals the famous KSS value with $\eta_\perp/s=1/4\pi$ (hereafter $\hbar=k_B=1$). Nevertheless, because of the non-zero Chern-Simons (CS) coupling, an analytical expression for the longitudinal shear viscosity is not currently accessible. Therefore, the $\eta_\parallel$ are computed numerically using the pseudo-spectrum method. Our findings reveal that the ratio $\eta_\parallel/s$ could exhibit non-monotonic behavior with respect to the magnetic field or temperature. In some cases, the ratio $\eta_\parallel/s$ decreases and then increases with respect to temperature, resulting in a distinct minimum at an intermediate temperature. This feature demonstrates a striking similarity to what is observed in most fluids, thereby paving the way for the construction of holographic models for strongly coupled fluids, such as quark-gluon plasma. Furthermore, at extremely low temperatures, the ratio $\eta_\parallel/s$ is finite for $B<B_c$ but approaches zero for $B>B_c$, suggesting its potential role as a useful order parameter for the QPT in present theory. On the other hand, the butterfly velocity has been computed using two independent methods: the shock wave analysis and the pole-skipping phenomenon. Both methods yield consistent results and correctly reproduce the splitting of butterfly velocities in the direction parallel to the magnetic field. Moreover, our study shows that each butterfly velocity displays an abrupt change across the QPT, demonstrating its ability to pinpoint the location of QCP through its derivative with respect to the magnetic field, \emph{i.e.} $\partial v_B/\partial B$. We also investigate the relation between shear viscosity and butterfly velocity.

The paper is organized as follows. In Section~\ref{sec2} we introduce the holographic model that exhibits the magnetic field-induced QPT. We study the behaviors of shear viscosity and butterfly velocity of the system in Section~\ref{sec3} and Section~\ref{sec4}, respectively. We conclude and discuss the open questions in Section~\ref{sec5}. More technical details are provided in the appendices.

\section{Holographic setup}\label{sec2}
We consider the five dimensional EMCS theory that is derived from a consistent truncation of Type IIB supergravity or M-theory\cite{Buchel:2006gb,Gauntlett:2006ai}
\bea\label{eq:action}
S=\frac{1}{16\pi G_N} \int d^5x \sqrt{-g}\Big( R+\frac{12}{L^2}-\frac{1}{4}F_{ab}F^{ab}+\frac{k}{24} \epsilon^{abcde}A_a F_{bc} F_{de} \Big) \,,
\eea
where the Maxwell field strength $F_{ab}=\partial_a A_b-\partial_b A_a$. Here $G_N$ is Newton's constant, $L$ is the AdS radius, and $k=\frac{2}{\sqrt{3} }$ is the CS coupling.\footnote{From the bottom-up point of view, the CS coupling $k$ can be treated as a free parameter. For $k>k_c\approx 1.158$, this system is known to be unstable below a critical temperature towards the formation of a helical order~\cite{Nakamura:2009tf,Donos:2012wi}.} The corresponding equations of motion read
\bea
G_{ab}-6g_{ab}-\frac{1}{2}\left(F_{ac}F_b^{\ c}-\frac{1}{4}g_{ab}F_{cd}F^{cd} \right) &=&0 \,,\\
\nabla_{b}F^{ba} +\frac{k}{8} \epsilon^{abcde}F_{bc}F_{de}&=&0 \,,
\eea
where we have set $16\pi G_N=1$ and $L=1$, without loss of generality.

To study the magnetic field-driven QPT, we consider the following dyonic black brane solution~\cite{DHoker:2010onp,Ammon:2016szz}:
\begin{equation}\label{eq:ansata1}
\begin{split}
&ds^2=\frac{1}{r^2}\Big[- \big(f-h^2p^2\big)dt^2+2ph^2 dtdz+g\big( dx^2+dy^2 \big)+h^2 dz^2+\frac{dr^2}{f}\Big]\,, \\
&A=A_t dt-\frac{B}{2}y dx+\frac{B}{2}x dy-A_z dz \,,
\end{split}
\end{equation}
where $f,\, g,\,h,\,p,\,A_t$ and $A_z$ are functions of the holographic coordinate $r$. The system approaches the asymptotically AdS boundary as $r\to0$, while it goes to the event horizon when $r\to r_h$. The temperature and entropy density of the system read
\bea
T=-\frac{f'(r)}{4\pi}\Big|_{r=r_h} \,,\quad s=\frac{4\pi g(r)h(r)}{r^3} \Big|_{r=r_h} \,.
\eea
Note that the magnetic field $B$ breaks Lorentz invariance along the $z$-axis, leading to spatial anisotropy of the system. 

After solving the bulk equations of motion, we can extract thermodynamic quantities, such as the free energy density, chemical potential and energy density. We point out that the CS term results in a non-trivial correction to the thermodynamics of the above dyonic black brane. In particular, the free energy density derived through the quantum statistical relation violates the standard form of the first law of thermodynamics. This issue has been addressed in our recent work~\cite{Cai:2024tyv} (see also Appendix~\ref{app:a} for more details). We shall work in the grand canonical ensemble by fixing the chemical potential $\mu=1$.

\subsection{Quantum criticality}

We now turn to the quantum criticality associated with the QPT induced by the magnetic field. When $B=0$, the system can be solved analytically and the solution is known as the electrically charged Reissner-Nordstr\"om (RN) black brane 
\bea
f=1-\left( 1+\frac{\mu^2r_h^2}{3} \right) \frac{r^4}{r_h^4}+\frac{\mu^2}{3r_h^4}r^6 \,,\quad A_t=\mu \left(1-\frac{r^2}{r_h^2} \right) \,,
\eea
with $g=h=1\,, p=A_z=0$.
As $T\to0$, the near horizon geometry of the RN solution is $AdS_2\times R^3$. On the other hand, for sufficiently strong magnetic field ($\vec{B}\propto \vec{z}$), the $x-y$ plane will stop to contract in the deep infrared region, resulting in $AdS_{3}\times R^2$ near horizon geometry~\cite{DHoker:2009mmn}. Consequently, the interplay between the charge density and the magnetic field will give rise to a QPT at the critical magnetic field $B_c\approx 0.332$~\cite{DHoker:2010onp}. In this case, numerical methods are necessary to obtain the full solution, for which we will employ the spectral method to solve the equations of motion. 

\begin{figure}[h]
\begin{center}
\includegraphics[width=0.68\textwidth]{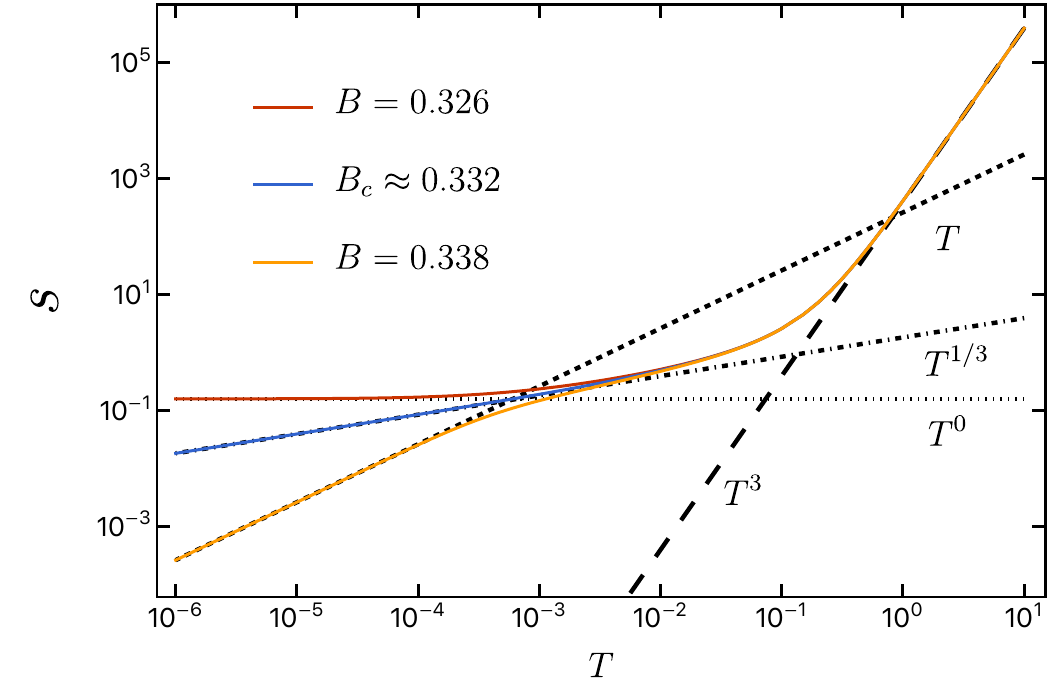}
\end{center}
\vspace{-0.3cm}
\caption{\small The entropy density $s$ as a function of temperature $T$ in different phases. There is a QPT at the critical magnetic field at $B_c\approx 0.332$. We consider $k=2/\sqrt{3}$ and fix the chemical potential $\mu=1$.}
\label{fig:critical-B}
\end{figure}

It has been established that scaling of entropy density near the QCP takes \cite{DHoker:2010zpp,DHoker:2010onp}
\bea
s=T^{1/3} \phi_{s} \left( \frac{B-B_c}{T^{2/3}} \right)\,,\label{eq:escaling}
\eea
where $\phi_s$ is a scaling function of the normalized magnetic field $B$ and temperature $T$. Following the standard scaling hypothesis, one can deduce from~\eqref{eq:escaling} that the dynamical critical exponent $z=3$ and the correlation length exponent $\nu=1/2$. As a demonstration, Figure~\ref{fig:critical-B} illustrates the temperature scaling of the entropy density at different phases. Above the quantum critical point \emph{i.e.} at $B=B_c\approx 0.332$, the entropy density scales as $s\sim T^{1/3}$, which confirms the analytical result~\eqref{eq:escaling}. When $B=0.326<B_c$, the entropy decreases to a constant in the extremal low temperature. In contrast, when $B=0.338>B_c$, the entropy displays a linear temperature dependence $s\sim T$ as $T\to0$. 

Within the Landau paradigm, phases of matter are characterized by their symmetries, and whether or not those are spontaneously broken which is associated with a local order parameter. The QPT defies the Landau paradigm as there is no symmetry breaking for all phases of~\eqref{eq:ansata1}. From the gravitational perspective, the mechanism driving this QPT can be interpreted as the expulsion of electric charge from the black hole horizon into the bulk, occurring as the magnetic field gradually increases from zero. This charge expulsion process terminates at a critical magnetic field strength $B=B_c$, where the black hole horizon becomes electrically neutral. Based on this charge expulsion mechanism, this QPT could be referred to as a transition from a ``fractionalized" phase with charged horizons ($B<B_c$) to a ``cohesive" phase with charged matter in the bulk ($B>B_c$)~\cite{DHoker:2012rlj} (see also~\cite{Hartnoll:2011pp,Hartnoll:2012ux}).

In the following part, we shall study the properties of shear viscosity and the butterfly velocity of the strongly coupled quantum many-body system dual to the EMCS theory. We are primarily concerned with how the presence of quantum criticality influences the behavior of shear viscosities and butterfly velocities.

\section{Shear viscosity}\label{sec3}
The viscosity tensor can be defined via the Kubo formula as
\bea
\eta_{ij,kl}= \lim_{\omega\to0} \frac{1}{\omega} \text{Im}[G^R_{ij,kl} (\omega, 0)] \,,
\eea
where the retarded correlation for the energy-momentum tensor is given by
\bea
G^R_{ij,kl}(\omega,0)=-\int dtd^3x e^{i\omega t} \theta(t) \langle [T_{ij}(t,\vec{x}), T_{kl}(0,0)]\rangle \,,
\eea
evaluated at zero wave-vector $\vec{k}=0$ and finite frequency $\omega$. The above retarded Green functions are holographically dual to the metric fluctuations on the background solutions. Therefore, all we need is to solve the bulk equations of motion for metric fluctuations which could be, in principle, coupled with other perturbations.

In our system, the spatial $SO(3)$ rotation symmetry is broken explicitly due to the presence of background magnetic field, but fortunately a $SO(2)$ symmetry along the transverse direction is preserved. For this specific symmetry breaking, there exist two shear viscosities which are related to the symmetric part of $\eta_{ij,kl}$, \emph{i.e.}
\bea
\eta_\parallel=\eta_{xz,xz}=\eta_{yz,yz}\,, \quad \eta_\perp=\eta_{xy,xy} \,.
\eea
Additionally, the presence of chiral anomaly contributes two more dissipationless Hall viscosities (antisymmetric part of $\eta_{ij,kl}$), \emph{i.e.}
\bea
\eta_{H\parallel}=\eta_{yz,xz}\,, \quad \eta_{H\perp}=\eta_{xy,xx}=\eta_{xy,yy} \,,
\eea
and we will not consider the Hall viscosities in the present work.

To determine the transports of the system, we should consider fluctuations of $g_{ab}$ and $A_b$ above the background solution (\ref{eq:ansata1}). The general form of the fluctuations takes
\bea
g_{ab}=\bar{g}_{ab}+ \int d\omega d^3 k e^{i k_{\mu} x^{\mu}} h_{ab}(r,k^{\mu}) \,,\quad A_{b}=\bar{A}_{b}+ \int d\omega d^3 k e^{i k_{\mu} x^{\mu}} a_{b}(r,k^{\mu}) \,,
\eea
where $\bar{g}_{ab}$ and $\bar{A}_b$ denote the background solutions, and $k^\mu=(\omega, \vec{k})$.
By choosing the radial gauge $a_r=0 ,h_{ra}=0$, we can classify the physical fluctuations according to how they transfer under the $SO(2)$ along the $(x,y)$ plane, see Table~\ref{table}.\,\footnote{One can find that $\eta_{H\perp}=0$ due to the decoupling between $h_{xy}$ and $h_{xx}-h_{yy}$.}

\begin{table}[htbp]  
\centering 
\begin{tabular}{c|c}  
Helicity & Fluctuations \\ 
\hline 
2 &   $h_{xy}\,, h_{xx}-h_{yy}$ \\
\hline
1 &   $h_{tx}\,, h_{ty}\,, h_{xz}\,, h_{yz}\,, a_x \,, a_y $  \\ 
\hline 
0 &  $h_{tt} \,, h_{tz} \,, h_{zz} \,, h_{xx}+h_{yy} \,, a_t \,, a_z $ \\
\end{tabular} 
\caption{Classification of fluctuations according to the helicity under $SO(2)$ rotation.} \label{table}
\end{table}

One can show that the helicity two perturbation $h_{xy}$ leads to the universal $\eta_{\bot}/s=1/4 \pi$ behavior, see Appendix~\ref{app:eta} for more details. On the other hand, the modes from the helicity one sector are responsible for a non-universal shear viscosity $\eta_\parallel$. Therefore, in our analysis, we will focus exclusively on the helicity one sector.

The computation of the non-universal shear viscosity is highly non-trivial, as these modes in the helicity one sector are all coupled to each other. It has been shown that for an anisotropic bulk system there could be an analytical horizon formula for the longitudinal shear viscosity, see \emph{e.g.}~\cite{Baggioli:2023yvc}. Unfortunately, all known examples were found for the static case, while the present geometry~\eqref{eq:ansata1} is not a static one due to the CS coupling. Therefore, we decide to obtain $\eta_\parallel$ by solving the equations of motion numerically. For numerical convenience, we introduce  
\bea
Y_{\pm}=h^x_{\ z} \pm i h^y_{\ z} \,,\quad  Z_{\pm}=h^x_{\ t} \pm i h^y_{\ t} \,,\quad  a_{\pm}=a_x \pm i a_y \,, \label{eq:perts}
\eea
that can split the helicity one fluctuations into two decoupled sectors: $(Y_{+}, Z_{+}, a_{+})$ and $(Y_{-}, Z_{-}, a_{-})$. Consequently, the simplified equations of motion for helicity one fluctuations can be written as
\bea
Y_{\pm}''+R_1 Y_{\pm}' +\left(  \frac{\omega^2}{f^2}-\frac{B^2 r^2}{f g^2}  \right) Y_{\pm}+\frac{h^2 p'}{f} Z_{\pm}' -\frac{r^2 A_z'}{g} a_{\pm}' &=&0 ,\, \\
Z_{\pm}''+S_1 Z_{\pm}' -\frac{B^2 r^2}{f g^2} Z_{\pm}+S_2 Y_{\pm}'+ \frac{\omega^2 p}{f^2} Y_{\pm}+\frac{r^2}{g}\Big( A_t' a_{\pm}' \mp \frac{\omega B}{f g} a_{\pm} \Big) &=&0,\,  \\
a_{\pm}''+ T_1 a_{\pm}'+ \left(\frac{\omega^2}{f^2}\mp \frac{k \omega r A_z'}{f h}\right) a_{\pm} + T_2 Z_{\pm}'+ T_3 Y_{\pm}' && \nonumber\\
 -\frac{B k r }{f h} \left( A_z' Z_{\pm}+ A_t' Y_{\pm} \right) \pm \frac{\omega B}{f^2} \left( Z_{\pm} -p Y_{\pm} \right) &=&0 \,,
\eea
together with the coefficients
\bea
R_1(r)&=&  \frac{f'}{f}+\frac{2g'}{g}-\frac{h'}{h}-\frac{h^2 pp'}{f}-\frac{3}{r}  \,,\nonumber \\ 
S_1(r)&=& \frac{2g'}{g}+\frac{h'}{h}+\frac{h^2 pp'}{f}-\frac{3}{r}   \,,\quad\quad
S_2(r)= p\left( \frac{f'}{f}-\frac{2h'}{h} \right) -p' -\frac{h^2 p^2 p'}{f}  \,,\nonumber\\
T_1(r)&=& \frac{f'}{f}+\frac{h'}{h}-\frac{1}{r} \,, \quad\quad\quad
T_2(r)=\frac{g}{f}\left(A_t'+pA_z'\right) \,,\quad\quad\quad T_3(u)=\frac{g A_z'}{h^2} -p T_2(r) \,. \nonumber
\eea
Since we are interested in the shear viscosity, we will only turn on the source term for $Y_{\pm}$, for which the equations can be solved numerically after imposing the ingoing boundary conditions at the event horizon $r=r_h$. The UV expansions for the fluctuations can be found in Appendix~\ref{app:a}. 

\begin{figure}[h]
\begin{center}
\includegraphics[width=0.7\textwidth]{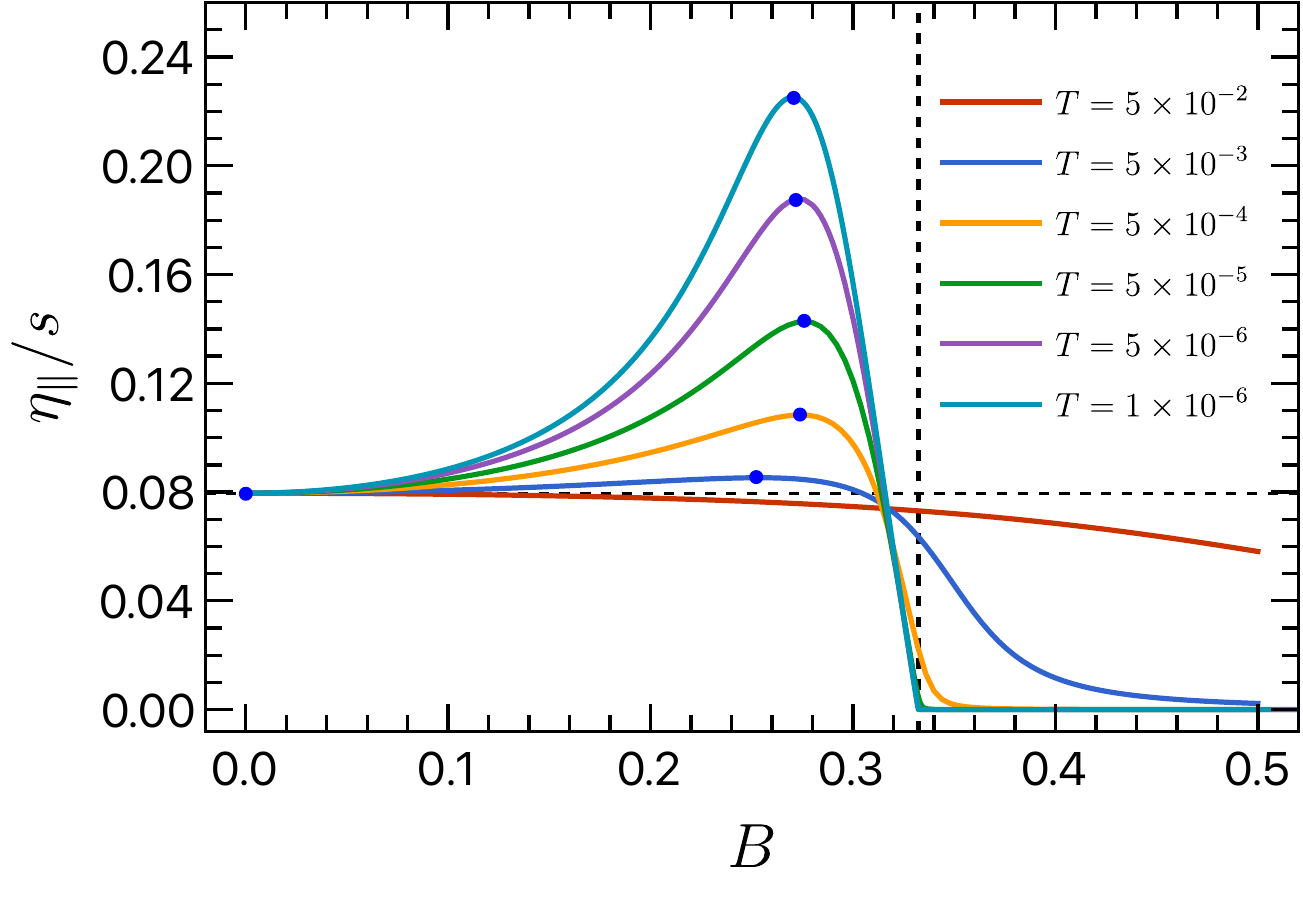}\quad
\end{center}
\caption{\small The ratio $\eta_{||}/s$ as a function of magnetic field at different temperatures. The blue point marks the maximum of each curve. The dashed horizontal line represents the KSS bound, while the dashed vertical line marks the location of QCP. We choose $k=2/\sqrt{3}$ and $\mu=1$.}\label{fig:etaB}
\end{figure}

\begin{figure}[h]
\begin{center}
\includegraphics[width=0.45\textwidth]{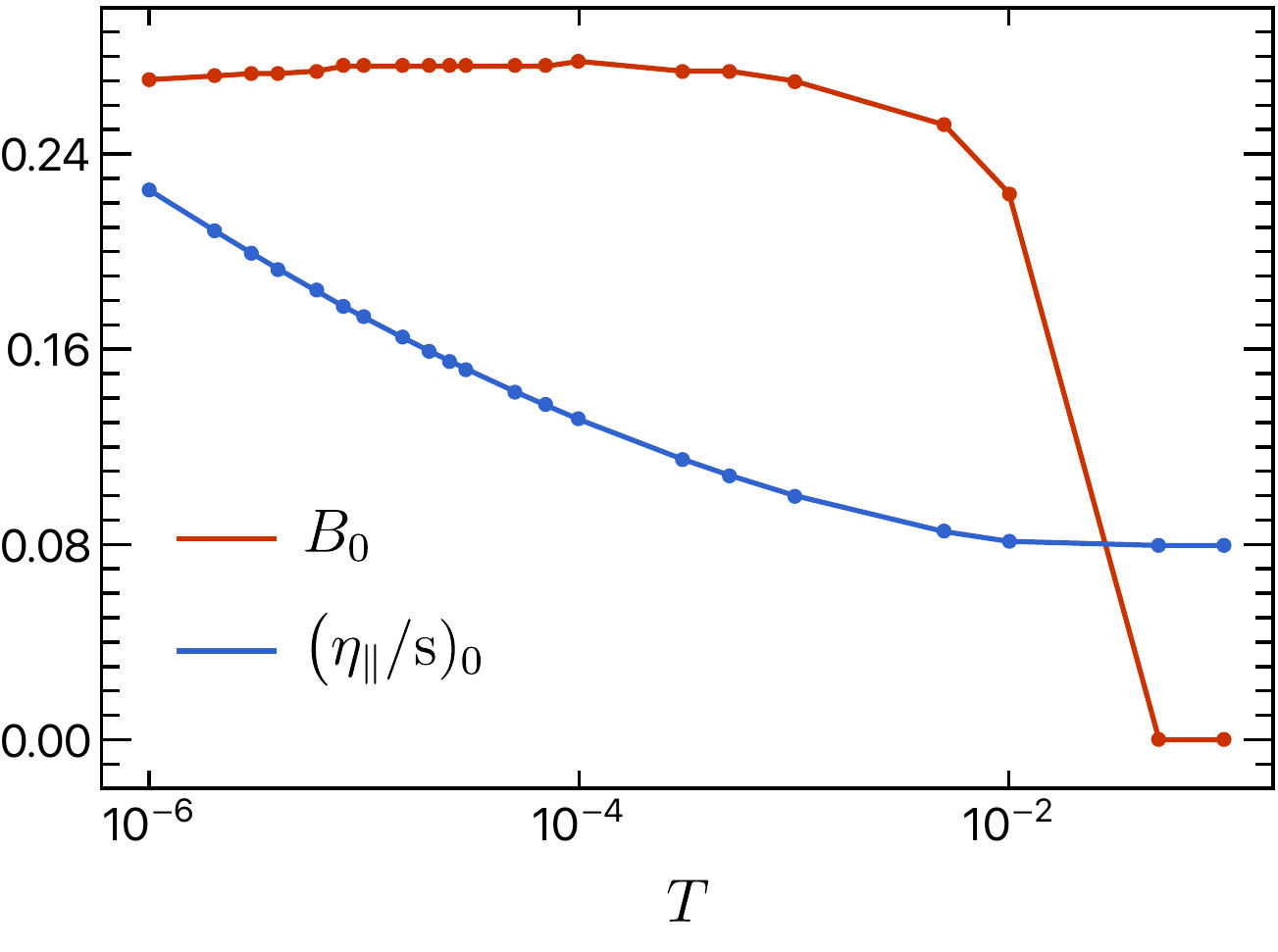}\quad
\includegraphics[width=0.49\textwidth]{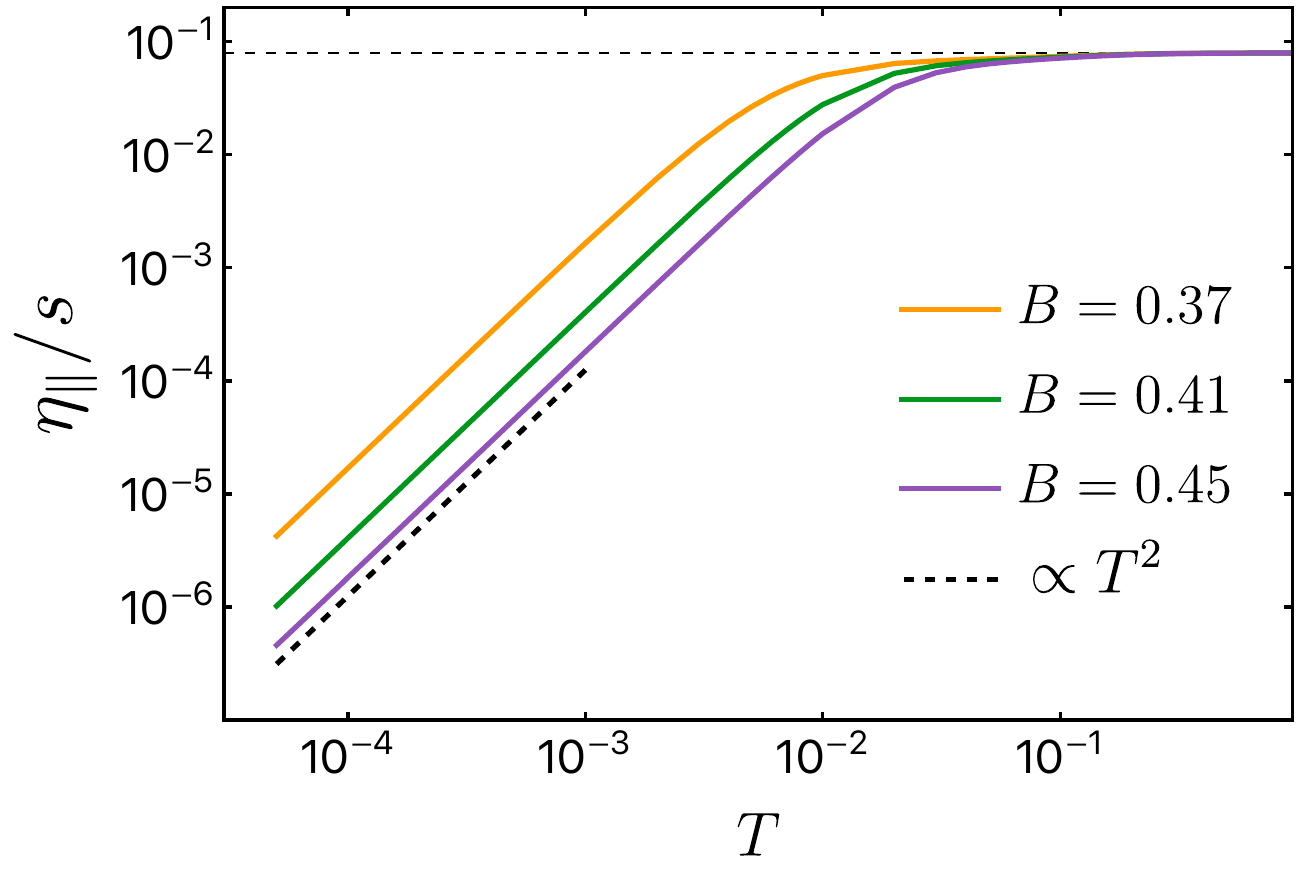}
\end{center}
\vspace{-0.3cm}
\caption{\small Left panel: The maximum $(B_0,(\eta_\parallel/s)_0)$ (blue points in Figure~\ref{fig:etaB}) as a functions of temperature. Right panel: The ratio $\eta_{||}/s$ as a function temperature for different $B>B_c$. The plots are generated for $k=2/\sqrt{3}$ and $\mu=1$.}\label{fig:etaBT}
\end{figure}

The behavior of $\eta_\parallel/s$ as a function of magnetic field at different temperatures is shown in Figure~\ref{fig:etaB}. In high temperature, starting from the KSS bound, the value of $\eta_\parallel/s$ decreases monotonically to a small value as the strength of the magnetic field increases (see the red curve). Interestingly, at low temperatures, the ratio $\eta_\parallel/s$ first increases to a maximum value $ (\eta_\parallel/s)_0$ at $B_0$ before the critical magnetic field $B_c$ (vertical line) is reached and then decreases monotonically to an extremely small value when $B>B_c$~\footnote{The behavior of $\eta_\parallel/s$ with respect to magnetic field was first studied in~\cite{Ammon:2020rvg}, yet working at a relative high temperature and strong magnetic field regime, \emph{i.e.} for $0.08<T/\mu<4$ and $0<B/\mu^2<18$. The non-monotonic magnetic field dependence of $\eta_\parallel/s$ there is parametrically small, and the crucial role played by the quantum criticality has not been addressed. As shown in Figure~\ref{fig:critical-B}, the quantum critical regime can only be touched at temperatures much lower than $T/\mu\sim 10^{-3}$.}. The location of the peak $(B_0, (\eta_\parallel/s)_0)$ that is marked by a blue point depends on the temperature.

In the left panel of Figure~\ref{fig:etaBT}, we show the behavior of $B_0$ and $(\eta_\parallel/s)_0$ with respect to temperature. While the value of $B_0$ (red curve) saturates at the low temperature limit, the maximum value of $\eta_\parallel/s$ (blue curve) increases monotonically as the temperature decreases. For the temperature range we show, the maximum value of $\eta_\parallel/s$ can be three times larger than the famous KSS value. In contrast, when $B>B_c$, the $\eta_\parallel/s$ ratio generally vanishes at zero temperature following a universal power-law $\sim T^2$ consistent with the result of $\eta_\parallel/s$ in purely magnetic black brane~\cite{Critelli:2014kra}, see the right panel of Figure~\ref{fig:etaBT}. In particular, for $B>B_c$, the entropy density scales as $s\propto T$ (see Figure~\ref{fig:critical-B}), indicating $\eta_\parallel =0$ as $T\to0$. Therefore, in the zero temperature limit $T\rightarrow 0$, we have
\begin{equation}
 \eta_\parallel/s>0\;\; (B<B_c), \quad\quad  \eta_\parallel/s=0\; \;(B>B_c)\,.
\end{equation}
This suggests that the low temperature limit of $\eta_\parallel/s$ could be a good order parameter for probing the QPT in the present system.\footnote{Although a good order parameter is absent at finite temperatures, one could identify an order parameter at zero temperature limit. For example, in the classical J-current model, the squared winding numbers serve as the order parameter and are adopted to identify the QCP through their scaling behavior near the QPT~\cite{Chen:2013ppa}.}

\begin{figure}[h]
\center
\subfigure{
\begin{minipage}[c]{0.485\textwidth}
\centering
\includegraphics[width=\textwidth]{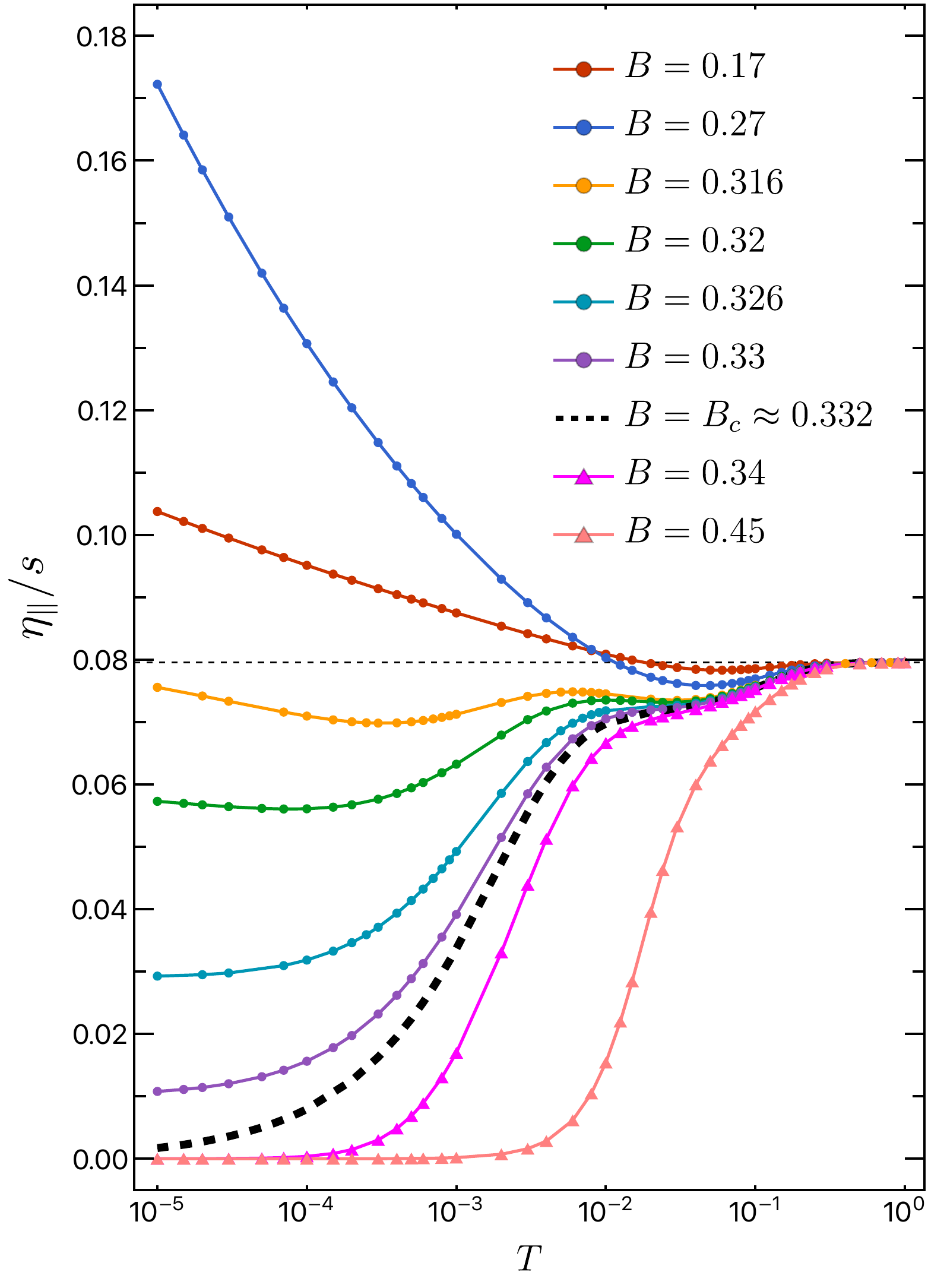}
\end{minipage}}
\subfigure{
\begin{minipage}[c]{0.48\textwidth}
\centering
\includegraphics[width=\textwidth]{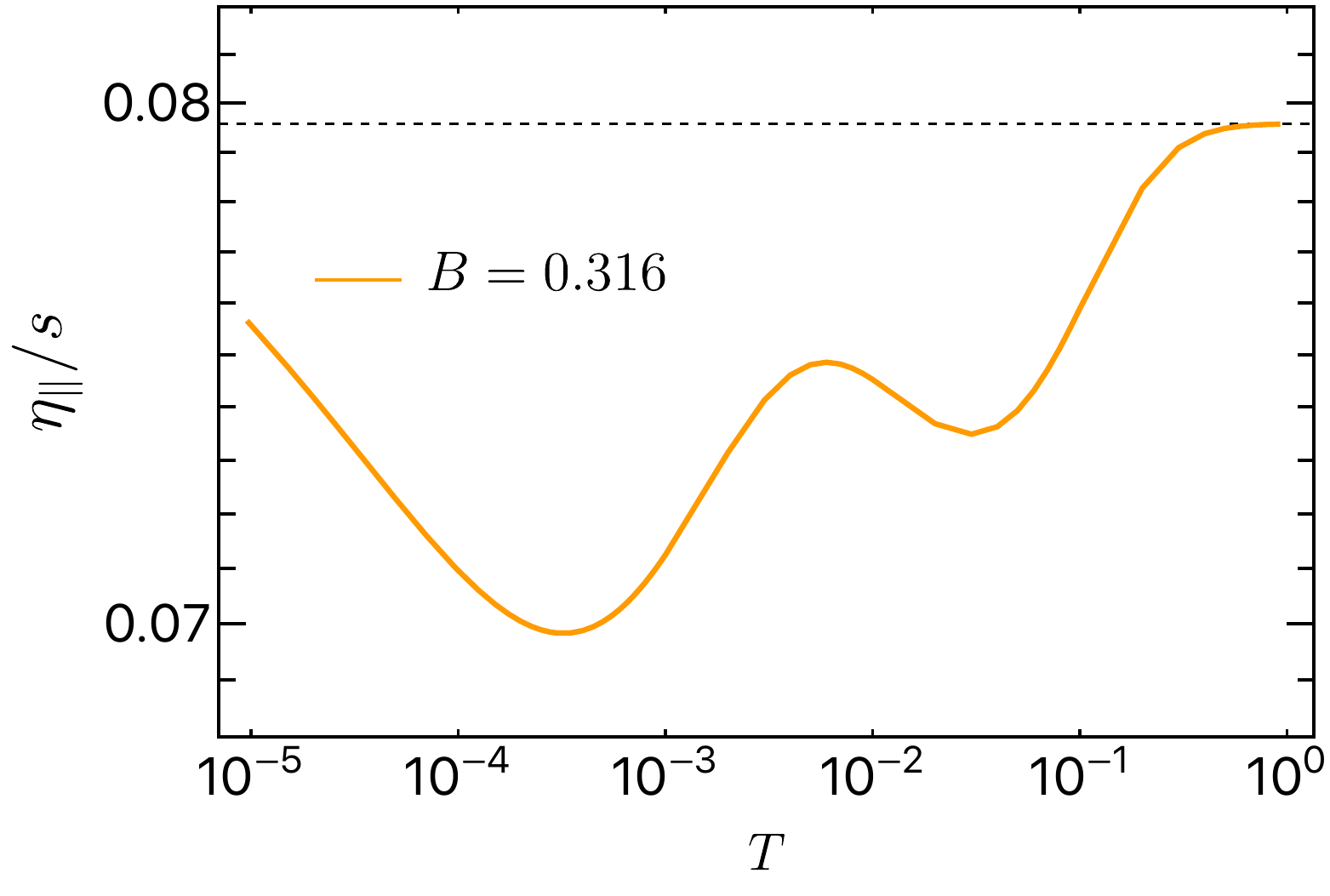}\vspace{4pt}
\includegraphics[width=\textwidth]{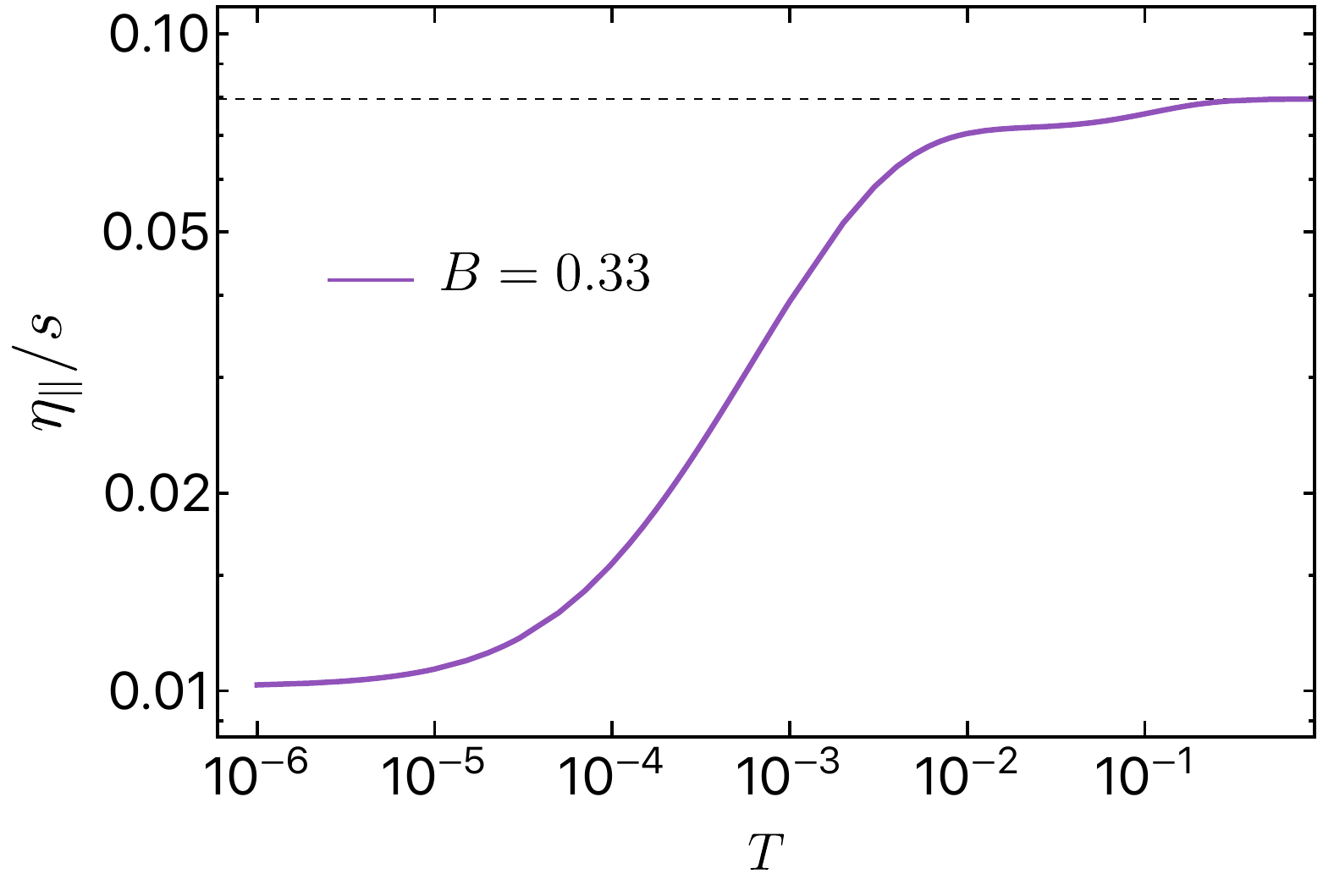}
\end{minipage}
}
\caption{Left panel: The ratio $\eta_{||}/s$ as a function of temperature for selected values of $B$. Right panel: $\eta_\parallel/s$ versus temperature at $B=B_c$ (top-right) and $B=0.33$ (bottom-right). The dashed horizontal line represents the KSS value $\eta/s=\frac{1}{4\pi}$. The plots are generated for $k=2/\sqrt{3}$ and $\mu=1$.}
\label{fig:etaT}
\end{figure}
Figure~\ref{fig:etaT} shows the temperature dependence of the ratio $\eta_\parallel/s$ in different magnetic fields. For small $B$, the ratio $\eta_\parallel/s$ first decreases and then increases to a value much higher than the value of the KSS bound as we decrease the temperature, see \emph{e.g.} the blue curve with $B=0.27$. Notably, to our knowledge, this should be the first holographic example that yields non-monotonic temperature dependence of $\eta_\parallel/s$ in the context of Einstein gravity\,\footnote{For a discussion on the temperature dependence of the shear viscosity to entropy ratio in the presence of higher derivative corrections, see~\cite{Cremonini:2011ej,Cremonini:2012ny}.}. More interestingly, when $B$ is relatively large, \emph{i.e.} $B>0.312$, the non-monotonic behavior in $\eta_\parallel/s$ persists, but it develops two local minima, see \emph{e.g.} the yellow curve with $B=0.316$. In this case, as the temperature drops, $\eta_\parallel/s$ first decreases, then increases, then decreases again, and finally increases (see the top-right panel for more details). It is worth to note that the ratio $\eta_\parallel/s$ exhibits a smooth minimum – distinct from the sharp change of shear viscosity at a phase transition – consistent with the absence of a finite-temperature phase transition in our holographic model~\cite{DHoker:2010zpp,Bernhard:2019bmu}.
However, the non-monotonicity disappears for a sufficiently strong magnetic field above a special value $B_t\approx0.975 B_c\approx 0.324$. For $B\in(B_t, B_c)$, $\eta_\parallel/s$ decreases monotonically to a small but finite value as the temperature drops, see \emph{e.g.} the purple curve with $B=0.33$ (also shown in an enlarged version in the bottom-right panel of Figure~\ref{fig:etaT}). For $B>B_c$, it vanishes at zero temperature following a power law $\sim T^2$, as demonstrated in the right panel of Figure~\ref{fig:etaBT}.

\begin{figure}[h]
\begin{center}
\includegraphics[width=0.65\textwidth]{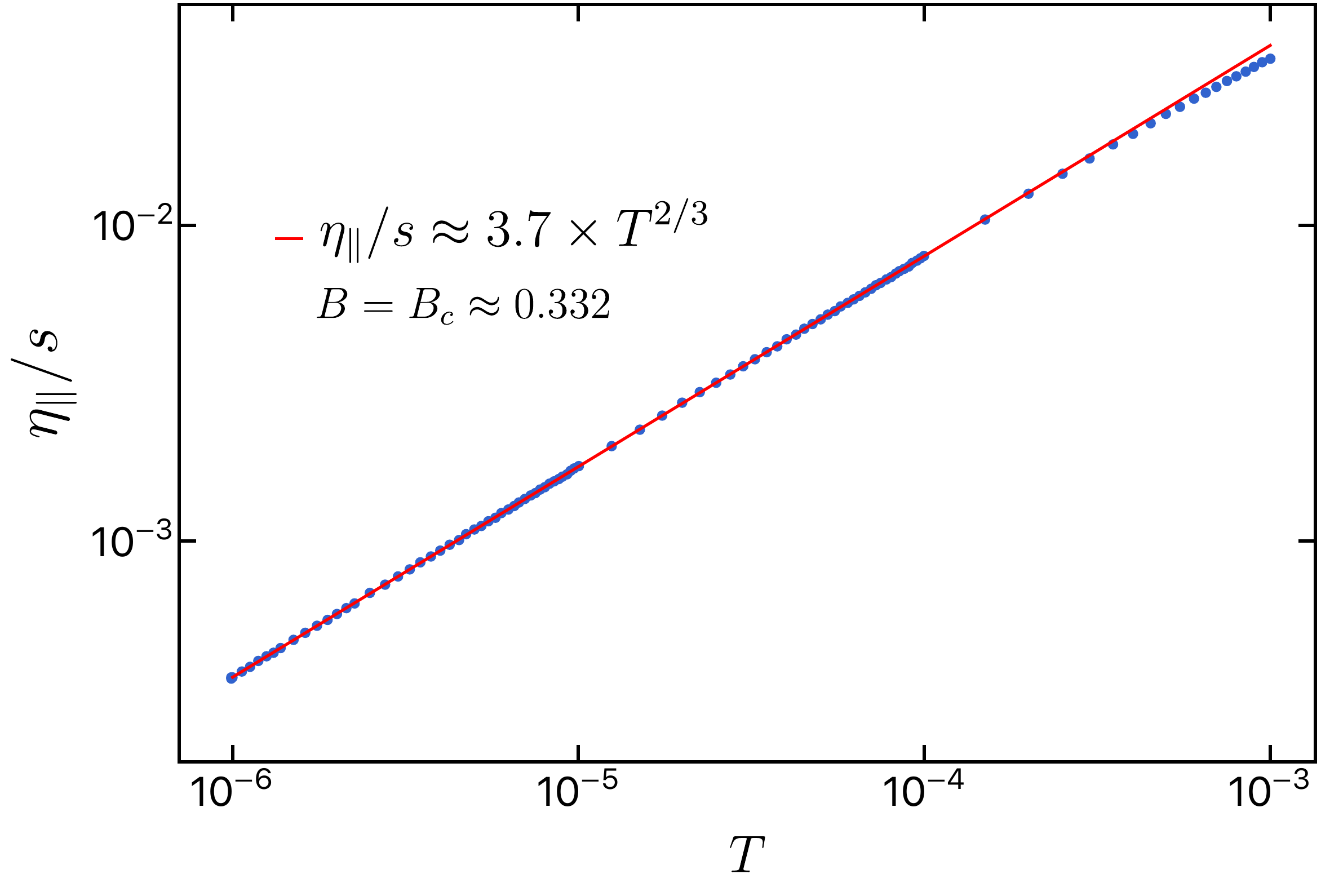}
\end{center}
\caption{\small Temperature dependence of $\eta_\parallel/s$ at $B_c$. The red solid curve is obtained by fitting low temperature data (denoted by blue dots). We have set $k=2/\sqrt{3}$ and $\mu=1$.}\label{fig:etaQCP}
\end{figure}

Of particular interest is the scaling behavior of $\eta_\parallel/s$ at $B_c$. 
As shown in Figure~\ref{fig:etaQCP}, the ratio $\eta_\parallel/s$ scales as
\bea\label{etaBc}
\eta_\parallel/s \propto T^{2/3}\,,
\eea
which is in sharp contrast to the quadratic scaling when $B>B_c$. The exponent $2/3$ herein can be understood as a direct combination of the dynamical critical exponent $z=3$ and the correlation length exponent $\nu=1/2$ of the QCP, \emph{i.e.} $2/3=1/(z\nu)$, see~\eqref{eq:escaling}. Moreover, the deviation from the scaling law~\eqref{etaBc} as increasing the temperature suggests that the quantum critical regime develops well below $T\sim 10^{-3}$. Nevertheless, the non-monotonic behavior of $\eta_\parallel/s$ versus $T$ for $B<B_t$ typically happens at  temperatures higher than $T\sim 10^{-3}$ (see the left panel of Figure~\ref{fig:etaT}). Moreover, the non-monotonic behavior of Figure~\ref{fig:etaT} disappears by turning off the CS coupling, see Appendix~\ref{app:c}.

\section{Butterfly velocity}\label{sec4}
In this section, we study the butterfly velocity using shock wave analysis as well as the phenomenon of pole-skipping. Our primary interest here is to carefully investigate the behavior of the butterfly velocity during QPT and to explore its relationship with the shear viscosity studied in the previous section.

\subsection{Butterfly velocity from shock wave geometry}
The sensitivity to initial conditions or chaos of the boundary field theory could be studied holographically through shock wave solutions~\cite{Shenker:2013pqa}. To study the shock wave on an eternal black hole background, it is more convenient to work with the smooth Kruskal coordinates $U$ and $V$, following~\cite{Abbasi:2019rhy,Roberts:2014isa}. More precisely, we consider the following coordinate transformation:
\bea
\begin{cases}
t-r^\ast=\hat{v} \\
t+r^\ast=\hat{u} 
\end{cases}
\Longrightarrow \quad
\begin{cases}
     V=e^{ -\frac{f'(r_h)}{2} \hat{v} }\\
     U=-e^{\frac{f'(r_h)}{2} \hat{u} } 
\end{cases}
\eea
with $UV=-e^{ f'(r_h) r^\ast} \,, U/V=-e^{f'(r_h) t}$ and $dr^\ast=\frac{dr}{ \sqrt{f(f-h^2p^2)} }$ the tortoise coordinate. The background ansatz~\eqref{eq:ansata1} is then rewritten as
\bea
\begin{split}
ds^2&=\frac{1}{r^2}\bigg[ -C dUdV+g\big( dx^2+dy^2 \big)+h^2 dz^2+\frac{2ph^2}{f'(r_h)} \left(  \frac{dU}{U}-\frac{dV}{V} \right) dz\bigg]\,, \label{eq:5Dkruskal} \\
A&=\frac{ A_t }{f'(r_h)} \left( \frac{dU}{U}-\frac{dV}{V} \right) -\frac{B}{2}y dx+\frac{B}{2}x dy-A_z dz \,,
\end{split}
\eea
where $C(U,V)=\frac{4(f-h^2p^2 )}{f'(r_h)^2 }e^{-f'(r_h)r^\ast }$.

Consider a few particles falling into the black hole from the left boundary of the eternal black hole. The effect of the perturbation is negligible if the particles are released at a small $t_w$ in the past. However, if $t_w$ becomes sufficiently large, the particle's energy, when measured at the $t=0$ slice, will grow exponentially, resulting in a significant backreaction on the background geometry. The shock wave geometry produced by these perturbations is described by (\ref{eq:5Dkruskal}) for $U>0$ and by (\ref{eq:5Dkruskal}) with a shift along $V$ direction $V\to V+ H(\vec{x})$ for $U<0$. After a redefinition of $V$ coordinate, the shock wave configuration is given by
\bea
\begin{split}
ds^2&= -\frac{C}{r^2} dUdV+\frac{g}{r^2}\big( dx^2+dy^2 \big)+\frac{h^2}{r^2} dz^2 +\frac{2p h^2}{f'(r_h) r^2 } \left( \frac{dU}{U}-\frac{dV}{V} \right) dz \\
& +\frac{ C }{r^2} H(\vec{x})\delta(U) dU^2  +\frac{2p h^2}{f'(r_h) r^2 V } H(\vec{x}) \delta(U) dUdz \,,  \\
A&= \frac{A_t}{f'(r_h)} \left( \frac{dU}{U}-\frac{dV}{V} \right)-\frac{B}{2}y dx+\frac{B}{2}x dy -A_z dz +\frac{A_t}{f'(r_h)V} H(\vec{x})\delta(U) dU \,. 
\end{split}
\eea
The above geometry is uniquely characterized by the magnitude $H(\vec{x})$, which can be determined by the $UU$-component of the Einstein equation. Plugging the above ansatz into the Einstein and Maxwell equations and using the fact $U\delta'(U)=-\delta(U)$ and $U\delta(U)=0$, we derive the (shock wave) equation for $H(\vec{x})$ as
\bea
\left[ -\partial_x^2 -\partial_y^2 -\frac{g}{h^2} \partial_z^2 -g p'\partial_z +m_0^2 \right] H(\vec{x}) \propto \frac{16\pi G_N E_0 g}{ C } e^{2\pi T t_w} \delta(\vec{x}) \,,\label{eq:shockwaveH}
\eea
where $m_0^2= \frac{g}{4} \left( 24-A_t'^2 -h^2p'^2 \right) -\frac{B^2}{4g}$ and all the background fields herein are evaluated at the event horizon. Note that $E_0$ and $t_w$ represent the initial energy and time of the particles, respectively. Unlike the situations in static black holes, there is a linear derivative term $\partial_z$ presented in~\eqref{eq:shockwaveH}, resulting in the splitting of butterfly velocities in the direction $\vec{z}$, as will be shown later.

To solve~\eqref{eq:shockwaveH}, it is convenient to express $H(\vec{x})$ in momentum space as
\bea\label{eqH}
H(\vec{x})= \int_{-\infty}^{+\infty} H_0 e^{i k^i x_i} d^3k_i \,.
\eea
By substituting~\eqref{eqH} into~\eqref{eq:shockwaveH}, $H$ can be computed using contour integration. Especially, for the case where $\vec{k}$ is perpendicular to $\vec{B}$, the solution is given by
\bea
H(x_{\perp})=\int_{-\infty}^{+\infty} e^{i k x_{\perp} } \frac{ e^{2\pi T t_w} E_0 }{k^2 -m_0^2} dk \sim \frac{\pi E_0}{m_0} e^{2\pi T(t_w-t_\ast) -m_0|x|} \,.
\eea
Here, $t_\ast=\frac{1}{2\pi T} \ln{ \left(\frac{c_0}{G} \right) } \approx \frac{ \ln{N^2} }{2\pi T}$ with $c_0$ a constant. Compared to~\eqref{eq:otoc}, the butterfly velocity along the transverse direction reads
\bea
v_B^{\perp}=\frac{2\pi T}{m_0} \,. \label{eq:vbtrans}
\eea
On the other hand, for the case where $\vec{k}$ is parallel to $\vec{B}$, the solution of $H$ is
\bea
H(z)&&=\int_{-\infty}^{+\infty} e^{i k z } \frac{h^2}{g} \frac{ e^{2\pi T t_w} E_0 }{k^2 +i \frac{h^2 p'}{f'} k-\frac{h^2}{g} m_0^2} dk  \nonumber\\
&&\sim \begin{cases}
\frac{ 2\pi h E_0 }{ g \sqrt{24-A_t'^2-B^2/g^2} } e^{ 2\pi T(t_w-t_\ast) -k_+ z }  \,,\,\,\quad  z>0 \\
\frac{ 2\pi h E_0 }{ g \sqrt{24-A_t'^2-B^2/g^2} } e^{ 2\pi T(t_w-t_\ast) - k_- |z| } \,,\,\quad  z<0
\end{cases}
\eea
where $k_\pm=\frac{h_0}{2}(h_0 p_0 \pm\sqrt{24-A_{t0}^2-B^2/g_0^2})$ with $p_0=-p'(r_h), A_{t0}=-A_t'(r_h)$. Therefore, we have two butterfly velocities in the direction $\vec{k} \parallel \vec{B}$, with
\bea\label{eq:vb12}
v_{B\pm}^{\parallel} = \frac{2\pi T}{ k_\pm } \,,\label{eq:vblong}
\eea
where $\pm$ symbols represent the information spreading parallel and antiparallel to the magnetic field direction, respectively.

The key result of~\eqref{eq:vb12} is that there exist two distinct butterfly velocities when measured in a direction parallel to the magnetic field, as first found in~\cite{Abbasi:2019rhy} for $\mu\ll T$ and $B\ll T^2$. Additionally, the splitting of longitudinal butterfly velocities supports the idea that chiral anomaly can be macroscopically manifested through the lens of quantum chaos~\cite{Abbasi:2019rhy,Abbasi:2023myj}. Similar splitting of butterfly velocity also exist in rotating black holes \emph{e.g.} Kerr-AdS$_4$ and Myers-Perry-AdS$_5$ black hole, which are understood as perturbations ``upstream" and ``downstream" against the direction of rotation~\cite{Amano:2022mlu,Blake:2021hjj}. Remarkably, the same expressions for butterfly velocity (\ref{eq:vbtrans}) and (\ref{eq:vblong}) can be extracted by studying the pole-skipping phenomenon~\cite{Blake:2018leo}. In order to verify the above results obtained from the shock wave analysis, we will also calculate the butterfly velocity by studying the energy density dynamics of the system.

\subsection{Butterfly velocity from pole-skipping}

To study the pole-skipping phenomenon, we should consider the near horizon perturbations with ingoing boundary conditions. One elegant way to impose the ingoing boundary condition at the black hole horizon is to work with the Eddington-Finkelstein coordinates. The coordinates transformation are
\bea
dv=dt-\frac{d r}{f} \,,\quad  d\tilde{z}=dz+\frac{p}{f}d r \,,
\eea
and the corresponding background becomes\,\footnote{Strictly speaking, the ingoing null coordinate should be $\hat{v}$ with $d\hat{v}=dt-\frac{dr}{ \sqrt{f(f-h^2p^2)} }$. However, as one approaches the black hole horizon, the $v$ coordinate becomes null coordinate and thus is equal to $\hat{v}$ at the leading order in $(r_h-r)$. Therefore, the ingoing mode corresponds to $e^{-i\omega v} \simeq e^{-i\omega \left( t +\frac{ \ln{(r_h-r)} }{4\pi T} \right)} =e^{-i\omega t} (r_h-r)^{-\frac{i\omega}{4\pi T}} \propto e^{-i\omega t} f^{-\frac{i\omega}{4\pi T}} $. The coordinates~\eqref{eq:EFc} have also been used in the study of quasi-normal modes of this model~\cite{Ammon:2017ded}.}
\bea
\begin{split}
ds^2&=\frac{1}{r^2}\Big[- \big(f-h^2p^2\big)dv^2 -2dv dr+2ph^2 dvd\tilde{z}+g\big( dx^2+dy^2 \big)+h^2 d\tilde{z}^2\Big]\,, \\
A&=A_t dv -\frac{B}{2}y dx+\frac{B}{2}x dy -A_z d \tilde{z} \,, \label{eq:EFc}
\end{split}
\eea
where we have used the gauge symmetry to set $A_r=-\frac{A_t+pA_z}{f}$ in~\eqref{eq:EFc} to zero.

Consider the energy density perturbation $\delta g_{vv}(r,v,x_i)=\delta g_{vv}(r) e^{-i\omega v+i k_i x^i}$ together with other perturbations that couple with it. At the black hole event horizon, the equation of $\delta g_{vv}$ takes
\bea
\begin{split}
\bigg[g k_\parallel^2+h^2k_\perp^2+ i k_\parallel gh^2p' +\frac{Br^2h^2}{2g}+ \frac{gh^2}{2r^2} \left( r^4A_t'^2+r^2h^2p'^2-24 \right) \qquad &\\
+ \frac{(i\omega+f') h}{r} ( rg'h+ rgh' -3gh)  \bigg] \delta g_{vv} \qquad  & \\
+\left( \omega+\frac{i f'}{2} \right)\Big[ 2\left( k_\parallel g\delta g_{v\parallel} +k_\perp h^2\delta g_{v\perp} \right)+ \omega \left( g\delta g_{zz} +h^2\delta g_{xx} +h^2 \delta g_{yy} \right) \Big] \Bigg|_{r=r_h}=&0 \,.
\end{split}
\eea
Therefore, $\delta g_{vv}$ decouples from other perturbations at $\omega=\omega_p=-\frac{i f'}{2}|_{r_h}=i 2\pi T$. Moreover, the pole-skipping point is defined as the special point $(\omega=\omega_p, k=k_p)$ for which the coefficient of $\delta g_{vv}$ is equal to zero and thus the above equation holds automatically~\cite{Blake:2018leo}. The butterfly velocity is then given by $v_B =\omega_p/k_p$. For $\vec{k}\perp \vec{B}$, we obtain the butterfly velocity
\bea
v_{B}^{\perp}=\frac{ 4\pi T}{\sqrt{g_0 (24-A_{t0}^2-h_0^2 p_0^2)-B^2/g_0} } \,. \label{eq:vbpole1}
\eea
On the other hand, for $\vec{k}\parallel \vec{B}$, the butterfly velocity reads
\bea
v_{B\pm}^{\parallel}=\frac{4\pi T}{p_0 h_0^2 \pm h_0\sqrt{24-A_{t0}^2-B^2/g_0^2} } \,, \label{eq:vbpole2}
\eea
where $\pm$ denote the information spreading along $+\vec{B}$ or $-\vec{B}$ direction, respectively. The expressions~\eqref{eq:vbpole1} and~\eqref{eq:vbpole2} derived from pole skipping are in complete agreement with the results~\eqref{eq:vbtrans} and~\eqref{eq:vblong} from shock wave analysis.
\begin{figure}[h]
\begin{center}
\includegraphics[width=0.49\textwidth]{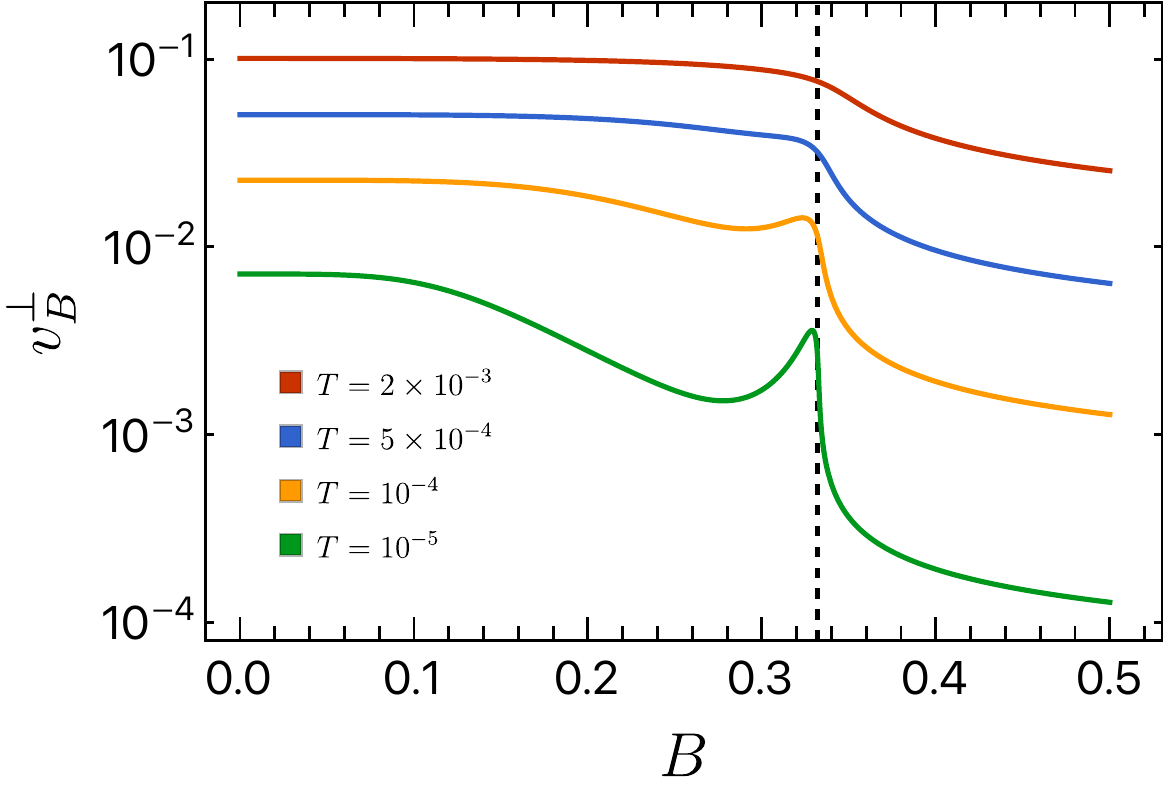}
\includegraphics[width=0.49\textwidth]{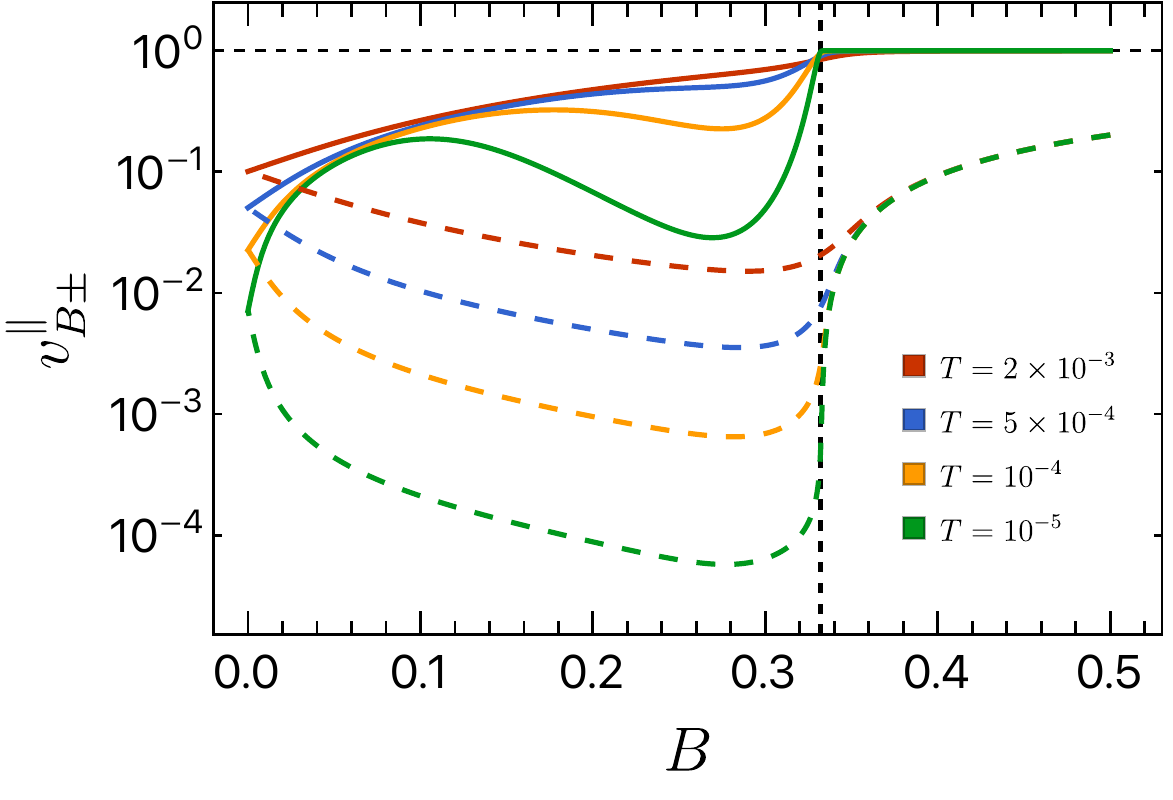}
\end{center}
\vspace{-0.3cm}
\caption{\small The three butterfly velocities ($v_B^\perp, v_{B\pm}^\parallel$) as a function of magnetic field at different temperatures. In the right panel, the solid lines represent the butterfly velocity $v^\parallel_{B+}$ while the dashed lines represent $v^\parallel_{B-}$. The location of $B_c$ is denoted by the dashed vertical line. The plots are generated for $k=2/\sqrt{3}$ and $\mu=1$.}
\label{fig:vB}
\end{figure}

Figure~\ref{fig:vB} shows the behavior of butterfly velocities as a function of background magnetic field at different temperatures. As the temperature drops, one finds that the overall amplitude of the transverse butterfly velocity $v^\perp_B$ (left panel) decreases, meanwhile a local maximum develops at $B_c$ for sufficiently low temperatures. On the other hand, the longitudinal butterfly velocities (right panel) always split into $v^\parallel_{B+}$ and $v^\parallel_{B-}$ and satisfy $v^\parallel_{B+} > v^\parallel_{B-}$, which was first found in~\cite{Abbasi:2019rhy,Abbasi:2023myj}. Above $B_c$, $v^\parallel_{B+}$ is always equal to $1$ and the difference between $v^\parallel_{B+}$ and $v^\parallel_{B-}$ becomes small by increasing $B$. The behavior of butterfly velocity (\emph{i.e.} $v^\parallel_{B+}=1$ for $B\geq B_c$) is notable, and it is helpful to obtain some intuitive explanations either from the low temperature behavior of~\eqref{eq:vbpole2} or even from the zero temperature infrared geometry. However, some difficulties arise in performing the zero-temperature analysis. First, it appears that the finite temperature analysis and the zero temperature limit do not commute and therefore may not explain each other~\cite{Natsuume:2020snz}. Moreover, the pole-skipping phenomena have been investigated in the zero-temperature, but its precise relation to the chaotic properties--particularly the butterfly velocity-- remains unclear~\cite{Natsuume:2020snz}. Second, the zero temperature geometry of the present holographic model is still under investigation though some near horizon solutions have been suggested~\cite{DHoker:2010onp}. Nonetheless, the special behavior of $v^\parallel_{B+}$ can be understood from the near horizon $BTZ\times R^2$ geometry induced by the strong magnetic field at sufficiently low temperatures, indicating that the dual system is effectively governed by a two-dimensional CFT dual to a BTZ black hole. This naturally explains the observed $v^\parallel_{B+}=1$ for $B \geq B_c$, which holds exactly for BTZ black hole~\cite{Jahnke:2019gxr,Mezei:2019dfv,Liu:2020yaf}.
In contrast, non-monotonic features are shown when $B<B_c$ at low temperatures. Particularly, approaching the limit $T\rightarrow 0$, one has $v^\parallel_{B+}=1$ and $v^\parallel_{B-}=0$ at $B_c$, a distinctive feature first observed in~\cite{Abbasi:2023myj}.\footnote{One can show that the denominator in~\eqref{eq:vbpole2} for $v^\parallel_{B-}$ remains finite as $T\to0$, indicating that $v^\parallel_{B-}=0$ at $B=B_c$.} These results imply that information propagates at the speed of light in the $+B$ direction when $B \geq B_c$, while it is effectively frozen in the $-B$ direction at the QCP.

\begin{figure}[htp]
\begin{center}
\includegraphics[width=0.99\textwidth]{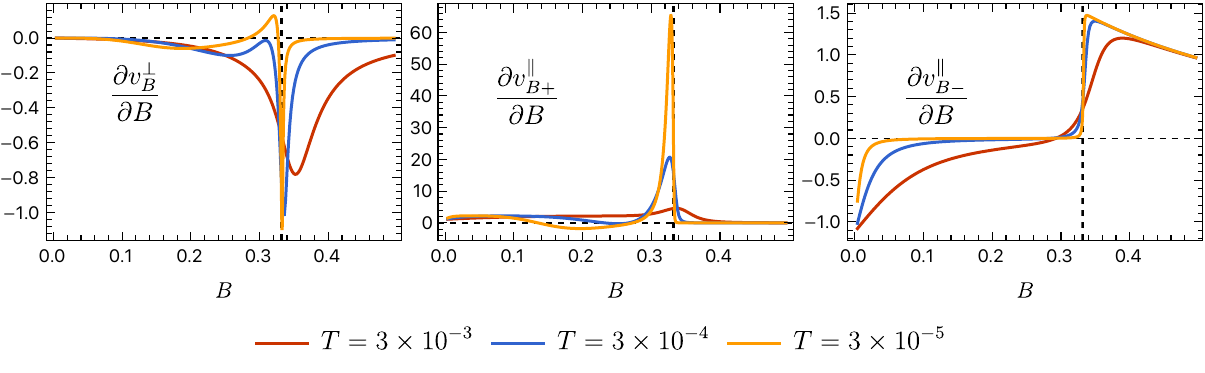}
\end{center}
\vspace{-0.3cm}
\caption{\small The derivative of butterfly velocities with respect to $B$ for different values of temperature. The vertical dashed line indicates the location of the QCP. Sudden changes are obvious near the QCP. We consider $k=2/\sqrt{3}$ and $\mu=1$.}
\label{fig:vBprime}
\end{figure}

The properties of longitudinal butterfly velocities $v^\parallel_{B\pm}$ have been suggested as an indicator of the QPT in~\cite{Abbasi:2023myj}. Nevertheless, in order to precisely locate the QCP, it is necessary to have two butterfly velocities readily available. Here, we demonstrate that each butterfly velocity can independently serve as indicators of the QPT, thanks to their abrupt changes in the vicinity of the QCP. More precisely, we focus on the derivative of the butterfly velocity with respect to $B$, which indicates the sensitivity of the butterfly velocity against the change in magnetic field. In Figure~\ref{fig:vBprime}, it is easy to see that $\partial v_B/\partial B$ shows a dip (left), a peak (middle) or a jump (right) as approaching the QCP. The drastic changes become more and more pronounced as we approach the QCP. On the other hand, thermal effects smooth out these changes, suggesting that $\partial v_B/\partial B$ might be good probes not only for the QCP but also for the quantum critical region.\footnote{Similar idea has been explored in holographic model with metal-insulator transition~\cite{Ling:2016ibq}.}

\subsection{Relation between shear viscosity and butterfly velocity}

Before ending this section, we consider the relation between shear viscosity and butterfly velocity. For many anisotropic but static systems, the butterfly velocities are related to the value of metric functions at the horizon. For a 5-dimensional static metric with
anisotropy in the $z$ direction, the horizon formula reads~\cite{Blake:2016wvh}\,\footnote{Note that for static case, there is no splitting of longitudinal butterfly velocities.}
\bea
\frac{(v^\parallel_B)^2}{ (v^\perp_B)^2} =\frac{g_{xx}}{g_{zz}} \Big|_{r\to r_h} \,.
\eea
Meanwhile, the shear viscosities satisfy
\bea
\frac{\eta_\parallel}{\eta_\perp} = \left( \frac{g_{xx}}{g_{zz}} \right)^{\mathcal{P}} \Big|_{r\to r_h} \,, \label{eq:horizon}
\eea
where $\eta_\perp/s=\frac{1}{4\pi}$. The exponent $\mathcal{P}=-1$ if the anisotropy is induced by the magnetic field~\cite{Critelli:2014kra} and $\mathcal{P}=1$ if the anisotropy is triggered by other fields, \emph{e.g.} axion field~\cite{Rebhan:2011vd} and vector hair~\cite{Baggioli:2023yvc}. This observation motivates us to conjecture a universal relation between shear viscosity and butterfly velocity, \emph{i.e.}
\bea
\frac{\eta_\parallel}{s} \left( \frac{ (v^\perp_B)^2 }{ (v^\parallel_{B})^2 } \right)^\mathcal{P} =\frac{1}{4\pi} \label{eq:etaVb}
\eea
This could serve as a novel bound for the shear viscosity to entropy ratio, generalizing the KSS bound to anisotropic systems. In particular, obtaining an expression for the butterfly velocity is relatively straightforward. In fact, this bound is valid in various anisotropic systems with different matter contents, including~\cite{Critelli:2014kra,Finazzo:2016mhm,Baggioli:2023yvc,Landsteiner:2016stv,Gursoy:2020kjd,Giataganas:2017koz}.

The model studied in this paper is anisotropic due to the presence of magnetic field, which motivates us to explore such a relationship within our model and subsequently derive an expression for the ratio $\eta_\parallel/s$. However, we numerically check that the new bound~\eqref{eq:etaVb} does not hold in the present case. The key difference compared to previous anisotropic models is that the bulk geometry of the EMCS model is stationary because of the non-trivial contribution from the CS coupling. In fact, the stationary properties and the quantum anomaly inherent in the system have resulted in the splitting of butterfly velocities in the direction parallel to the magnetic field and thus increase the difficulty in constructing such a bound.

Nonetheless, we numerically check the temperature dependence of the butterfly velocities at $B=B_c$ and find that
\bea
v^\perp_{B} \propto T^{\alpha}  \,,\quad  v^\parallel_{B-} \propto T^{\alpha} \,,\quad  1-v^\parallel_{B+} \propto T^\beta  \,,
\eea
where $\alpha=0.666\approx 2/3$ and $\beta=0.469$. This implies that the temperature scaling of $v^\perp_{B}$ or $v^\parallel_{B-}$ has the same exponent as that of the ratio $\eta_\parallel/s$ above the QCP. Although~\eqref{eq:etaVb} is violated for the stationary case, it is possible to obtain a new bound for the stationary case by generalizing~\eqref{eq:etaVb} with an  appropriate combination of longitudinal and transverse butterfly velocities, \emph{e.g.}
\bea
\frac{\eta_\parallel}{s} \left( \frac{ (v^\perp_B)^2 }{ \mathcal{F}( v^\parallel_{B+}, v^\parallel_{B-})^2 } \right)^\mathcal{P} =\frac{1}{4\pi} \,, \label{eq:etaVb1}
\eea
where $\mathcal{F}$ is defined as a function of $v^\parallel_{B+}$ and $v^\parallel_{B-}$. For the static case where $v^\parallel_{B+}=v^\parallel_{B-}=v^\parallel_{B}$, we should have $\mathcal{F}=v^\parallel_{B}$ and~\eqref{eq:etaVb1} recovers~\eqref{eq:etaVb}.

\section{Conclusion and discussion}
\label{sec5}

We have studied the shear viscosity and butterfly velocity in a class of strongly coupled quantum many body systems holographically dual to the EMCS theory, which allows a magnetic field driven QPT with the QCP at $B_c$. Our work is a natural continuation of~\cite{Ammon:2020rvg,Abbasi:2019rhy,Abbasi:2023myj} which initiated this line of research. Since no symmetry is broken in this QPT, there exists no conventional order parameter. Nevertheless, from the bulk point of view, it can be considered as a transition from a ``fractionalized" phase with charged horizons ($B<B_c$) to a ``cohesive" phase with charged matter ($B>B_c$).

In addition to confirming previous results in the literature, we have uncovered some interesting and novel features. First, we found that the longitudinal shear viscosity-entropy ratio $\eta_\parallel/s$ versus temperature shows a very rich behavior, depending on the value of $B$, see Figure~\ref{fig:etaT}. For a relatively small magnetic field, $\eta_\parallel/s$ initially decreases as the temperature is lowered and rises to a value much larger than the KSS bound. This creates a minimum for $\eta_\parallel/s$ at an intermediate temperature, which resembles the minimum observed in the liquid-gas transition. Specifically, this minimum in $\eta_\parallel/s$ emerges smoothly in the intermediate temperature range. This behavior closely parallels the shear viscosity observed during gas-like to liquid-like transitions in water or helium at supercritical regimes (see Figure 1 in~\cite{Bernhard:2019bmu}). Moreover, our results have revealed multiple minima in the temperature dependence of $\eta_\parallel/s$, indicating the complexity and richness of shear viscosity for magnetic field-induced QPT~\cite{DHoker:2010onp,DHoker:2012rlj}. It would be fascinating to explore whether such multiple minima occur in correlated quantum matters~\emph{e.g.}~\cite{Rost:2009,Tokiwa:2016,Naritsuka:2023}.
This is the first explicit example that yields non-monotonic temperature dependence of $\eta/s$ in Einstein gravity. Although the reason for such a minimum is still unclear, we note that this non-monotonic feature develops at relatively high temperatures and thus might be not closely related to the ground state geometry and the QCP. Even more interesting is the fact that the QPT is visible in the low-temperature behavior of the shear viscosity. Upon approaching QCP, $\eta_\parallel/s$ is finite for $B<B_c$ and vanishes for $B>B_c$. The vanishingly small of $\eta_\parallel/s$ at $B>B_c$ implies that one may understand the system as developing a layered structure with nearly frictionless flow between different layers. Therefore, the value of $\eta_\parallel/s$ can serve as the order parameter that distinguishes the two states of the system and identifies the QCP through their low temperature behaviors. What's more, we elaborated further on the fact that the splitting of longitudinal butterfly velocities at $T\rightarrow 0$ encodes important information about the QCP. More interestingly, we have uncovered that not only the longitudinal butterfly velocities but also the transverse butterfly can be used to identify the location of QCP. As shown in Figure~\ref{fig:vBprime}, all the butterfly velocities undergo rapid changes near the QCP, rendering the peak or singularity in their derivatives $\partial v_B/\partial B$ as reliable indicators of the QCP. We also discuss the possible relationship between shear viscosity and butterfly velocity.

There are many interesting questions that could be explored in future work. First, it would be interesting to derive an analytical expression for $\eta_{\parallel}/s$, perhaps by adopting the method outlined in \cite{Demircik:2023lsn,Donos:2022uea} or through some variation of the method used in \cite{Baggioli:2023yvc,Landsteiner:2016stv}. This endeavor will not only deepen our understanding of the non-trivial behavior of $\eta_\parallel/s$ but will also be invaluable in exploring the relationship between shear viscosity and other transport coefficients~\cite{Blake:2016wvh}. Second, we have limited ourselves to the holographic field theory with the supersymmetric value of the CS coupling. It would be interesting to study the shear viscosity and butterfly velocity for other values of CS coupling $k$, testing the universality of our findings. Third, we anticipate analogous non-monotonic temperature dependence of the shear viscosity to entropy ratio to emerge in other cases. For example, in holographic Weyl semimetals with momentum relaxation~\cite{Zhao:2021pih,Zhao:2021qfo}, the interplay between spontaneous symmetry breaking driven by a sourceless scalar field and anisotropy induced by the axial $U(1)$ gauge field could potentially give rise to a minimum in $\eta/s$.  Moreover, the critical theory in our present study is effectively 1+1 dimensional, because massless propagation at the critical point takes place along a single spatial direction. It is also interesting to study quantum criticality in higher dimensions. In particular, the case of 2+1 dimensions is relevant to various layered materials such as cuprates. Finally, it is of significance to test our results with some experimental setups.

\vspace{-.2cm}
\subsection*{Acknowledgments}
We thank Tan Chen, Hao-Tian Sun and Wei-Jia Li for useful discussions. This work was partly supported by the National Natural Science Foundation of China Grants No.\,12075298, No.\,12122513, No.\,12247156, and No.\,12047503. We acknowledge the use of the High Performance Cluster at the Institute of Theoretical Physics, Chinese Academy of Sciences.

\vspace{.2 cm}
\appendix
\section{Holographic renormalization}\label{app:a}
Close to the conformal boundary $r\to0$, the metric and gauge field take the following form:
\bea
\begin{split}
f(r)&= 1+2\xi r+\xi^2r^2+\frac{B^2}{6}r^4\ln{r}+f_4 r^4+\cdots \,, \\
g(r)&= 1+2\xi r+\xi^2r^2-\frac{B^2}{12}r^4\ln{r}-h_4 r^4+\cdots  \,, \\
h(r)&= 1+\xi r+\frac{B^2}{12}r^4\ln{r}+h_4 r^4+\cdots  \,, \\
p(r)&= p_4 r^4-4p_4\xi r^5+\cdots    \,, \\
A_t(r)&= \mu-\frac{\rho}{2} r^2+\xi\rho r^3+\cdots  \,, \\
A_z(r)&= A_{z2} r^2-2\xi A_{z2} r^3+\cdots\,.    
\end{split}
\eea
Note that, the constant $\xi$ reflects the reparameterization freedom $r\to r+\xi$ along the radial direction and can be set to zero. Near the event horizon $r=r_h$, the bulk fields behave as
\bea
\begin{split}
f(r)&= (r_h-r)f_0+\cdots  \,,\quad g(r)= g_0+\cdots  \,,\qquad\qquad  h(r)= h_0+\cdots  \,, \\
p(r)&= (r_h-r)p_0+\cdots    \,,\quad A_t(r)= (r_h-r)A_{t0}+\cdots \,,\quad A_z(r)= A_{z0}+\cdots \,.  
\end{split}
\eea
Using the scaling symmetry 
\begin{equation}
 \left(t,x,y,z,r \right)\to  c \left( t,x,y,z,r \right),\, \left(A_t, A_z \right)\to c^{-1}\left(A_t, A_z\right),\, B\to c^{-2}B ,\, (f,g,h,p)\to(f,g,h,p)\,,\nn  
\end{equation}
we can fix the location of the event horizon $r_h=1$.

Moreover, the UV expansion for the helicity one perturbations~\eqref{eq:perts} reads
\bea
\begin{split}
h^x_{\ z}&=h_{xz}^{(0)}+\frac{\omega^2 h_{xz}^{(0)}}{4} r^2+h_{xz}^{(1)} r^4+\Big( \frac{B^2}{4}-\frac{\omega^4}{16} \Big) h_{xz}^{(0)} r^4\ln{r}+\cdots \,, \\
h^y_{\ z}&=h_{yz}^{(0)}+\frac{\omega^2 h_{yz}^{(0)}}{4} r^2+h_{yz}^{(1)} r^4+\Big( \frac{B^2}{4}-\frac{\omega^4}{16} \Big) h_{yz}^{(0)} r^4\ln{r}+\cdots \,, \\
h^x_{\ t}&=h_{tx}^{(0)}+h_{tx}^{(1)} r^4+\Big( \frac{B^2 h_{tx}^{(0)}}{4}+\frac{i \omega B a_y^{(0)}}{4} \Big)  r^4\ln{r}+\cdots \,,\\
h^y_{\ t}&=h_{ty}^{(0)}+h_{ty}^{(1)} r^4+\Big( \frac{B^2 h_{ty}^{(0)}}{4}-\frac{i \omega B a_x^{(0)}}{4} \Big)  r^4\ln{r}+\cdots \,,\\
a_{x}&=a_{x}^{(0)}+a_x^{(1)} r^2-\Big( \frac{\omega^2 a_{x}^{(0)}}{2}+\frac{i\omega B h_{ty}^{(0)}}{2} \Big) r^2\ln{r}+\cdots \,, \\
a_{y}&=a_{y}^{(0)}+a_y^{(1)} r^2-\Big( \frac{\omega^2 a_{y}^{(0)}}{2}-\frac{i\omega B h_{tx}^{(0)}}{2} \Big) r^2\ln{r}+\cdots \,, 
\end{split}
\eea
where $h_{tx}^{(1)}$ and $h_{ty}^{(1)}$ are determined by $a_x^{(1)}, a_y^{(1)}$ as well as the leading order source terms.

The renormalised on-shell action is given by
\bea\label{eq:sren}
S_{ren}=S+S_{bdy}
\eea
where
\bea
S_{bdy}=\frac{1}{16\pi G}\int d^4x\sqrt{-\gamma}\Big[ 2K-\frac{6}{L}-\frac{L\hat{R}}{2}+\frac{L^3}{4}\ln{\left(\frac{r}{L}\right)} \left( \hat{R}_{\mu\nu}\hat{R}^{\mu\nu}-\frac{\hat{R}^2}{3}- \frac{\hat{F}^2}{L^2} \right) \Big].\nonumber
\eea
Here $\gamma_{\mu\nu}$ is the induced metric at the conformal boundary, $K$ is the trace of extrinsic curvature and $\hat{R}_{\mu\nu}$ denotes the Ricci tensor associated with $\gamma_{\mu\nu}$.

The stress tensor and current of the dual field theory are
\bea
\begin{split}
\langle T_{\mu\nu} \rangle &=  \lim_{r\to0} \frac{ L^2}{r^2} \Big[ -2K_{\mu\nu} +2(K-3)\gamma_{\mu\nu} +\hat{G}_{\mu\nu} +\ln{r} ( \hat{F}_{\mu\rho} \hat{F}_\nu^{\ \rho}-\frac{ \gamma_{\mu\nu} }{4} \hat{F}^2 -h_{\mu\nu}^{(4)}) \Big] \,, \\
\langle J^{\mu} \rangle  &= \lim_{r\to0} \frac{L^4}{r^4} \Big[ n_r \Big( F^{\mu r} +\frac{k}{6}\epsilon^{r\mu \alpha\beta\gamma} A_\alpha F_{\beta\gamma} \Big) +\nabla_\alpha \hat{F}^{\alpha\mu} \ln{r} \Big] \,, 
\end{split}
\eea
with
$h_{\mu\nu}^{(4)}= \hat{R}_{\mu\rho\nu\lambda}\hat{R}^{\rho\lambda} -\frac{1}{6}\hat{\nabla}_\mu \hat{\nabla}_\nu \hat{R}+\frac{1}{2}\hat{\nabla}^2 \hat{R}_{\mu\nu}-\frac{1}{3}\hat{R}\hat{R}_{\mu\nu}-\frac{\gamma_{\mu\nu}}{4}(\hat{R}_{\rho\lambda}\hat{R}^{\rho\lambda}-\frac{\hat{R}^2}{3}+\frac{\hat{\nabla}^2 \hat{R} }{3})$.

Therefore, the non-zero components of the stress tensor $\langle T_{\mu\nu}\rangle$ are
\bea
\begin{split}
\epsilon &= \langle T_{tt}\rangle=-3f_4\,, \,\,\qquad\qquad\qquad   \mathcal{P}_\perp=\langle T_{xx}\rangle=\langle T_{yy}\rangle=-\frac{B^2}{4}-f_4-4h_4\,,  \\ 
\mathcal{P}_\parallel  &= \langle T_{zz}\rangle=-f_4+8h_4 \,, \,\quad\qquad \langle T_{tz}\rangle=\langle T_{zt}\rangle=4p_4 = -\frac{k}{2}B\mu^2 \,,  \\
\langle T_{tx}\rangle&= \rho a_x^{(0)}+\frac{i\omega B a_y^{(0)} }{2}+\Big( \frac{B^2}{4} -f_4 -4h_4\Big)h_{tx}^{(0)} -\frac{2i B}{\omega}\Big( a_y^{(1)} -\frac{\rho h_{ty}^{(0)}}{2}+A_{z2} h_{yz}^{(0)} \Big)\,, \\ 
\langle T_{ty}\rangle&= \rho a_y^{(0)}-\frac{i\omega B a_x^{(0)} }{2}+\Big( \frac{B^2}{4} -f_4 -4h_4 \Big)h_{ty}^{(0)} +\frac{2i B}{\omega}\Big( a_x^{(1)} -\frac{\rho h_{tx}^{(0)}}{2}+A_{z2} h_{xz}^{(0)} \Big)\,, \\
\langle T_{xz}\rangle&= 4 h_{xz}^{(1)}-h_{xz}^{(0)} \Big( \frac{3\omega^4}{16}+ f_4 +4h_4 \Big)\,, \qquad
\langle T_{yz}\rangle = 4 h_{yz}^{(1)}-h_{yz}^{(0)} \Big( \frac{3\omega^4}{16}+ f_4 +4h_4 \Big)\,, 
\end{split}
\eea
while the non-zero components of the dual current $\langle J_\mu \rangle$ are
\bea
\begin{split}
\langle J_t\rangle &= -\rho \,,\qquad\quad  \langle J_z\rangle=-2A_{z2} =kB\mu \,, \\
\langle J_{x}\rangle&= 2a_x^{(1)} -2kB\mu h_{xz}^{(0)}-\frac{\omega}{2}\big(\omega a_x^{(0)}+iB h_{ty}^{(0)} \big)\,, \\ 
\langle J_{y}\rangle&= 2a_y^{(1)} -2kB\mu h_{yz}^{(0)}-\frac{\omega}{2}\big( \omega a_y^{(0)}-iB h_{tx}^{(0)} \big)\,.
\end{split}
\eea
where the covariant current is considered by dropping the contribution of the Chern-Simons term.

The free energy density $w$ can be derived from the Euclidean on-shell action, known as the \emph{quantum statistical relation}. One then obtains that
\bea
w &=& \frac{W}{V}=-\frac{S_{ren}}{V} = \epsilon -Ts-\mu \langle J^t \rangle-\frac{k B}{3}\int_0^{r_h} A_t\, A_z' d r \,, 
\eea
where the last term is from the CS coupling. It has been recently revealed that the above free energy density $w$ does not satisfy the first law of thermodynamics~\cite{Cai:2024tyv}, \emph{i.e.}
\bea
\delta w = -\left(s+ \frac{ k Q_{cs} }{T} \right)\delta T -\rho \delta\mu -k \delta Q_{cs} - M_B \delta B\,,
\eea
where 
\bea
Q_{cs} &=&\frac{B}{6}\int_0^{r_h} (A_z' A_t - A_z A_t') dr =\frac{B}{3}\int_0^{r_h} A_t A_z'dr \,,\nonumber\\
M_B &=& -\left( \int_0^{r_h} \left[ \frac{B}{r} \left( \frac{h}{g}-1 \right) +\frac{k}{2}(A_t A_z' - A_t' A_z) \right] dr +B\ln{r_h} \right) \,.\nonumber
\eea
This issue can be solved by adding the non-local term $Q_{cs}$ to the free energy~\cite{Cai:2024tyv}:
\bea
\tilde{w} =w +kQ_{cs} =\epsilon -Ts -\mu\langle J^t\rangle \,,
\eea
and one recovers the standard form of the first law of thermodynamics
\bea
\delta \tilde{w} =-s \delta T -\langle J^t \rangle \delta\mu -M_B\delta B \,,
\eea
where $M_B$ is recognized as the magnetization of the system.

\section{Transverse shear viscosity}\label{app:eta}
For transverse shear viscosity, we can directly obtain an analytical expression following the method used in~\cite{Baggioli:2023yvc} (see also~\cite{Landsteiner:2016stv,Baggioli:2018bfa}). Consider the fluctuation of the helicity two mode, denoted as $h_{xy}$. Defining $X=g^{xx}h_{xy}$, the equation of motion for $X$ reads
\bea
X''+\left( \frac{f'}{f}+\frac{g'}{g}+\frac{h'}{h}-\frac{3}{r} \right)X'+ \frac{\omega^2}{f^2} X=0 \,.\label{eq:spin2}
\eea
The UV expansion of $X$ is given by
\bea
X=X^{(0)}+\frac{\omega^2 X^{(0)}}{4}r^2+X^{(1)} r^4-\frac{\omega^4 X^{(0)} }{16} r^4\ln{r}+\cdots \,,
\eea
and the corresponding stress tensor reads
\bea
\langle T_{xy}\rangle &=& \left[ \langle T_{xx}\rangle +i\omega \eta \right]X^{(0)}= \left[ \langle T_{xx}\rangle+\frac{ 4X^{(1)} }{ X^{(0)} }-\frac{3\omega^4}{16} \right]X^{(0)}\,.
\eea

We then obtain the shear viscosity
\bea
\eta_{\bot}=\lim_{\omega\to0} \frac{1}{i \omega} \left( \frac{ 4X^{(1)} }{X^{(0)}}-\frac{3\omega^4 }{16} \right)=\lim_{\omega\to0} \frac{1}{i \omega} \frac{ 4X^{(1)} }{X^{(0)}} \,,
\eea
where $X^{(0)}$ can be normalized to one. Therefore, to obtain $\eta_\bot$, we only need to know $X^{(1)}$ at leading order in $\omega$, which can be expressed analytically by the horizon data. More precisely, we expand $X$ in powers of $\omega$,
\bea
X=&f^{-\frac{i\omega}{4\pi T} } \left( X_0+\omega\, X_1 +\cdots \right) \,,
\eea
where the temperature-dependent prefactor is to ensure that the perturbations obey the ingoing boundary conditions at the horizon. The equation~\eqref{eq:spin2} can be solved order by order in powers of $\omega$. We have
\bea
X_0=1 \,,\quad\quad  X_1=\int_1^r \left( \frac{i f'}{4\pi T f}+i g_0h_0\frac{\tilde{r}^3}{fgh} \right)d\tilde{r}\,.
\eea
Note that, $X_1$ is fixed by regularity condition near the black hole horizon. The near boundary expansion of $X$ at leading order in $\omega$ reads
\bea
X=1+\omega c_0+ \frac{i \omega g_0h_0}{4} r^4 +\cdots\,,
\eea
where $c_0=\int_1^0 \left( \frac{i f'}{4\pi T f}+i g_0h_0\frac{\tilde{r}^3}{fgh} \right)d\tilde{r}$ is a constant. Then, we can find that $X^{(0)}=1+\omega c_0$ and $X^{(1)}=i\omega g_0h_0/4$ for small $\omega$. Therefore, we get
\bea
\eta_{\bot}=\lim_{\omega\to0} \frac{1}{i \omega} \frac{ 4X^{(1)} }{X^{(0)}}=\lim_{\omega\to0} \frac{1}{i \omega} \frac{ i\omega g_0h_0 }{1+\omega c_0} =g_0h_0,\quad \Rightarrow\quad \frac{ \eta_{\bot} }{s}=\frac{1}{4\pi}\,.
\eea
%

\section{Longitudinal shear viscosity for $k=0$}\label{app:c}
To compare with the key results shown in the main text, we present additional results of the ratio of longitudinal shear viscosity to entropy density at zero CS coupling \emph{i.e.} $k=0$, for which the bulk spacetime~\eqref{eq:ansata1} becomes static with $p=A_z=0$. As shown in Figure~\ref{fig:etaT}, the intriguing non-monotonic temperature dependence of $\eta_\parallel/s$ emerges at small magnetic field in the presence of CS coupling. Consequently, our analysis will focus on the small $B$ regime.


\begin{figure}[h]
\begin{center}
\includegraphics[width=0.65\textwidth]{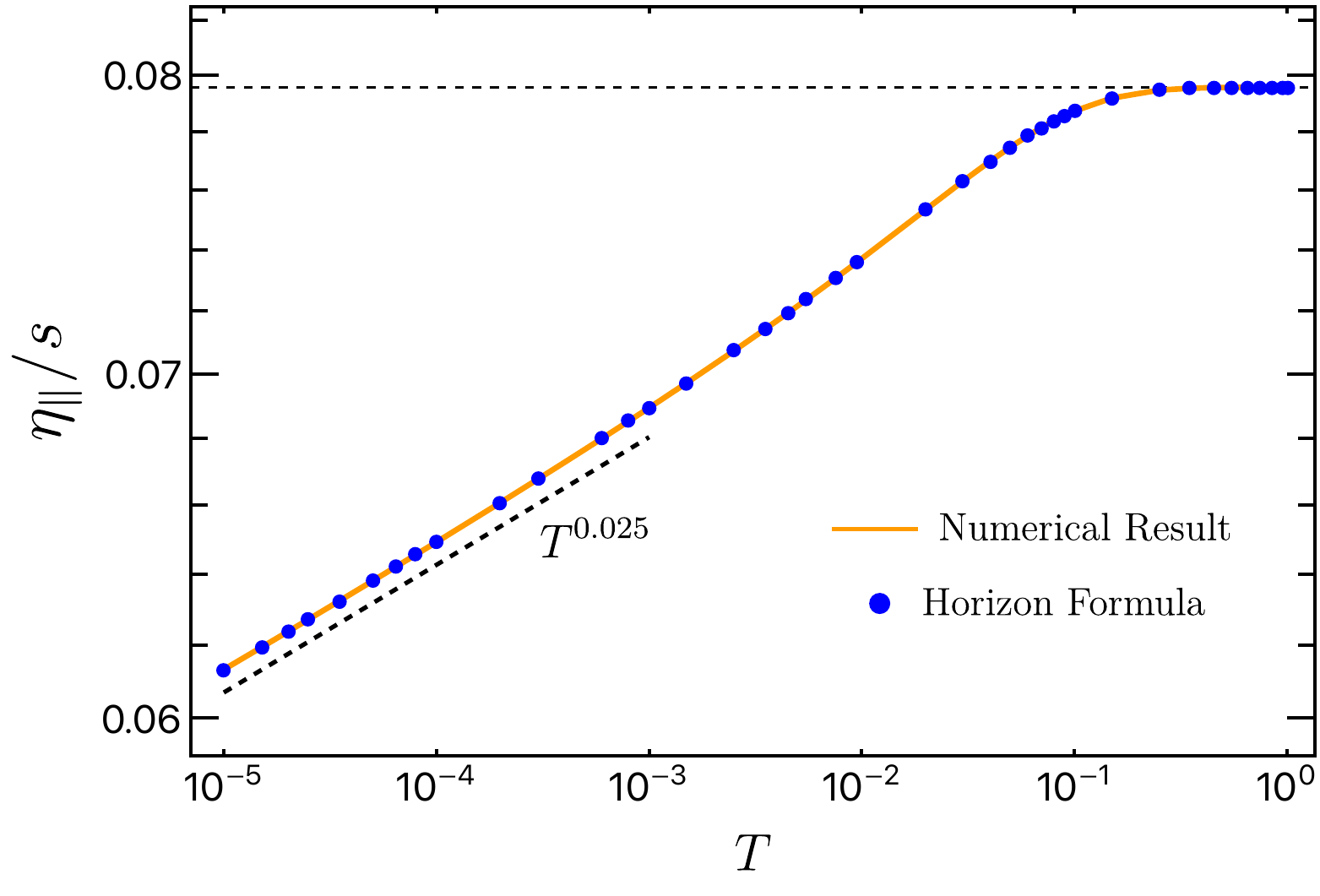}
\end{center}
\caption{\small The temperature dependence of $\eta_\parallel/s$ at $B=0.1$ for the CS coupling $k=0$. The solid line is from the numerical data and blue dots are from our horizon formula~\eqref{eq:horizon}. The black dashed curve is the low temperature scaling of $\eta_\parallel/s$ obtained by fitting the low temperature data. 
We have set $\mu=1$.}\label{fig:etak0}
\end{figure}

Figure~\ref{fig:etak0} shows the behavior of $\eta_\parallel/s$ as a function of temperature at $B=0.1$. The ratio $\eta_\parallel/s$ decreases monotonically as the temperature decreases, in contrast to the non-monotonic behavior revealed for $k=2/\sqrt{3}$ demonstrated in Figure~\ref{fig:etaT}. As $T\to0$, the $\eta_\parallel/s$ vanishes following a power law:
\bea
\eta_\parallel/s \propto T^{0.025} \,.
\eea
It is worth noting that the temperature scaling exponent observed here differs from that of pure magnetic black branes~\cite{Critelli:2014kra}. This discrepancy may be attributed to the interplay between charge density and magnetic field when $B$ is small. Furthermore, we check the numerical results with the horizon formula~\eqref{eq:horizon}, observing excellent agreement. As demonstrated in Figure~\ref{fig:etak0}, the numerical data (displayed with solid line) are in perfect agreement with the horizon formula~\eqref{eq:horizon}, shown with blue dots.


\end{document}